\pdfoutput=1
\documentclass[10pt,oneside,pdftex,dvipsnames,a4paper]{article}

\usepackage{etex}
\reserveinserts{18}

\usepackage{a4wide}
\usepackage{setspace,xspace,makeidx,fancyvrb,relsize}
\usepackage{alltt}

\usepackage{amsmath,amsthm,amssymb,amsfonts}
\usepackage{parallel,accents}
\usepackage{stmaryrd}

\usepackage{amsxtra,amstext,latexsym,dsfont} % for \mathds{N}
\normalfont
\usepackage{bm}

\usepackage[comma]{natbib}
\usepackage[symbol]{footmisc}
\usepackage{perpage}
\MakePerPage[2]{footnote}

\usepackage[pdftex]{graphicx}
\usepackage[dvipsnames]{xcolor}

\usepackage{pgf,tikz}
\usetikzlibrary{snakes}
\usetikzlibrary{backgrounds}
\usetikzlibrary{arrows}
\usetikzlibrary{patterns}
\usetikzlibrary{fit}
\usetikzlibrary{calc}

\usepackage{lscape} 
\usepackage{shorttoc}

\usepackage{array,longtable,booktabs,float,rotating}
\usepackage{dcolumn,ctable}

\usepackage[labelfont={bf},ruled,labelsep=period,small]{caption}

\usepackage[flushleft]{threeparttable}
\usepackage[linkcolor=black,colorlinks=true,citecolor=black,%ForestGreen,%pagebackref,%
bookmarks=true,bookmarksopen=true,bookmarksnumbered=true,bookmarksopenlevel=0,%
hyperindex=true,pdfpagemode=UseOutlines,linktocpage=true,pdfpagelayout=TwoColumnLeft,
pdftitle=Biostatistics,pdfstartview=FitH,pdffitwindow=true,urlcolor=blue]{hyperref}
\usepackage{ragged2e}

\usepackage{multirow,multicol}

\usepackage{fancyhdr,extramarks}

 \makeatletter%--------------------------
 \newcommand\sub{\@startsection%
     {subsubsection}{5}{0mm}{-1\baselineskip}{.01\baselineskip}%
     {\normalfont\itshape}}
 \renewcommand\subsubsection{\@startsection%
     {subsubsection}{3}{0mm}{-1\baselineskip}{.01\baselineskip}%
     {\normalfont\itshape}}
 \makeatother%--------------------------

\makeatletter
        \newcommand\Appendix[2][?]{%
            \refstepcounter{section}%
            \addcontentsline{toc}{appendix}%
                {\protect\numberline{\appendixname~\thesection}#1}%
            {\raggedleft\bfseries \appendixname\
                \thesection\par \centering#2\par}%
                \sectionmark{#1}%
                \@afterheading
                \addvspace{\baselineskip}}
        \newcommand\sAppendix[1]{%
            \raggedleft\bfseries\appendixname\par
            \@afterheading\addvspace{\baselineskip}}
    \makeatother

\normalsize \makeindex

\newcolumntype{A}{>{\centering}p{100pt}}
\newlength\savedwidth

\def\coldot{.}%
{\catcode`\.=\active%
    \gdef.{$\egroup\setbox2=\hbox to \dimen0 \bgroup$\coldot}}
\def\rightdots#1{%
    \setbox0=\hbox{$1$}\dimen0=#1\wd0%
    \setbox0=\hbox{$\coldot$}\advance\dimen0 \wd0%
    \setbox2=\hbox to \dimen0 {}%
    \setbox0=\hbox\bgroup\mathcode`\.="8000 $}
\def\endrightdots{$\hfil\egroup\box0\box2}

\newcolumntype{d}[1]{D{.}{.}{#1}}
\newcolumntype{A}{>{\centering}p{100pt}}
\newcolumntype{.}{D{.}{.}{-1}}
\newcolumntype{P}[2]{>{#1\raggedright\arraybackslash}p{#2}}

\DeclareFontFamily{U}{euc}{}% I chose euc because the chart is called Euler cursive
\DeclareFontShape{U}{euc}{m}{n}{<-6>eurm5<6-8>eurm7<8->eurm10}{}%
\theoremstyle{plain}      
\theoremstyle{plain}      
\theoremstyle{plain}      
\theoremstyle{plain}      
\theoremstyle{definition} 
\theoremstyle{definition} 
\theoremstyle{definition} \newtheorem{exa}{Example}
\theoremstyle{plain} 
\theoremstyle{definition} 
\theoremstyle{plain} \newtheorem{pro}{Proposition}
\theoremstyle{definition} 
\theoremstyle{definition} 
\theoremstyle{definition} 

\newcounter{nctr}
\usepackage[rightbars,color]{changebar}
\cbcolor{blue}
\VerbatimFootnotes
\usepackage{url}

\newcommand\tb{\textbf}
\newcommand\ti{\textit}

\newcommand\te{\text}
\newcommand\ma[1]{\te{\bf{#1}}}
\newcommand\ca{\mathcal}

\newcommand\op{\operatorname}

\newcommand\lb{\lbrace}
\newcommand\lt{\left}

\newcommand\qq{\qquad}

\newcommand\rb{\rbrace}
\newcommand\rt{\right}

\newcommand\tth{^\text{th}}

\newcommand\Bb{\ma{b}} %%% b
\newcommand\br{\ma{r}} % Vectors of neurones
\newcommand\bR{\ma{R}} % Correlation matrix
\newcommand\bW{\ma{W}} % Weight matrix
\newcommand\bX{\ma{X}}
\newcommand\bY{\ma{Y}}
\newcommand\bZ{\ma{Z}}

\newcommand\cE{\ca{E}} % Subset
\newcommand\cI{\ca{I}} % Fisher Information
\newcommand\cV{\ca{V}} % Volume
\newcommand\cW{\ca{W}} %
\newcommand\bbe{\bm\beta}

\newcommand\bep{\bm\epsilon}

\usepackage{fancyhdr}
\begin{document}
\sloppy
% Spacing.
%\onehalfspacing
%\doublespacing
%\setstretch{3} 
%\setcounter{secnumdepth}{-2}

%-----------------------------------------------
% title Page
\begin{center}
Running Head: \uppercase{Statistical Network Analysis for Functional MRI}
\end{center}
\vspace{3cm}
\begin{center}
\Large{\textbf{Statistical Network Analysis for Functional MRI: 
               \\ Summary Networks and Group Comparisons.}}
\\
\vspace{2.5cm} \normalsize
               Cedric E.~ Ginestet${^{abc}}$.
               Arnaud P.~ Fournel${^{d}}$, 
               and Andrew Simmons${^{bc}}$,
\end{center}
\begin{center}
\vspace{1cm}
  \rm ${^a}$Department of Mathematics and Statistics, Boston
  University, Boston, MA. \\
  \rm ${^b}$Department of Neuroimaging, Institute of Psychiatry,
  King's College London, UK. \\
  \rm ${^c}$National Institute of Health Research, Biomedical
  Research Centre for Mental Health and Biomedical Research Unit for
  Dementia, UK. \\
  \rm ${^d}$Laboratoire d'Etude des M\'{e}canismes Cognitifs (EMC),
  EA 3082, Universit\'{e} Lyon II, France. \\
\end{center}
%%%%%%%%%%%%%%%%%%%%%%%%%%%%%%%%%%%%%%%%%%%%%%%%%
\vspace{3cm}
\sub{Acknowledgments}
%\footnotesize
CEG's work was supported by a grant from the Air Force
Office for Scientific Research (AFOSR), whose grant number is
FA9550-12-1-0102; and by a fellowship from the UK National Institute
for Health Research (NIHR) Biomedical Research Center for Mental
Health (BRC-MH) and Biomedical Research Unit for Dementia at the South
London and Maudsley NHS Foundation Trust
and King's College London. CEG and AS were supported by the UK National Institute
for Health Research (NIHR) Biomedical Research Centre for Mental
Health (BRC-MH) at the South London and Maudsley NHS Foundation Trust
and King's College London and by the Guy's and St Thomas' Charitable
Foundation as well as the South London and Maudsley Trustees. 
APF has received financial support from the Region
Rh\^{o}ne-Alpes and the Universit\'{e} Lumi\`{e}re Lyon 2 through an
Explora’Doc grant. 
% We would also like to thank one anonymous reviewer for his/her
% valuable inputs.

%\vspace{1/2cm}
\sub{Correspondence}
Correspondence concerning this article should be sent to Cedric
Ginestet at the Department of Mathematics and Statistics, 
College of Arts and Sciences, Boston University, 111 Cummington Mall,
Boston, MA 02215. Email may be sent to \rm cgineste@bu.edu.
\pagebreak
%-----------------------------------------------
%%%%%%%%%%%%%%%%%%%%%%%%%%%%%%%%%%
\begin{abstract}
Comparing networks in neuroscience is hard, because the
topological properties of a given
network are necessarily dependent on the number of edges in that network. This
problem arises in the analysis of both weighted and unweighted
networks. The term density is often used in this context, in order to
refer to the mean edge weight of a weighted network, or to the number of
edges in an unweighted one. Comparing families of networks is therefore
statistically difficult because differences in topology are
necessarily associated with differences in density. 
In this review paper, we consider this problem from two different
perspectives, which include (i) the construction of summary
networks, such as how to compute and visualize the summary network from a
sample of network-valued data points; and (ii) how to test for
topological differences, when two families of networks also
exhibit significant differences in density. In the first instance,
we show that the issue of summarizing a family of networks can be
conducted by either adopting a mass-univariate approach, which
produces a statistical parametric network (SPN).
In the second part of this review, we then highlight the inherent problems
associated with the comparison of topological functions of families of networks
that differ in density. In particular, we show that a wide range of
topological summaries, such as global efficiency and network modularity are
highly sensitive to differences in density. Moreover, these problems
are not restricted to unweighted metrics, as we demonstrate that the
same issues remain present when considering the weighted versions of
these metrics. We conclude by encouraging caution, when reporting such
statistical comparisons, and by emphasizing the importance of constructing summary
networks.
\end{abstract}
%%%%%%%%%%%%%%%%%%%%%%%%%%%%%%%%%%
KEYWORDS: Networks, N-back, Statistical Parametric
Network (SPN), Small-world topology, 
Working memory, Weighted density, Density-integrated metrics.
%%%%%%%%%%%%%%%%%%%%%%%%%%%%%%%%%%

%%%%%%%%%%%%%%%%%%%%%%%%%%%%%%%%%%
%%%%%%%%%%%%%%%%%%%%%%%%%%%%%%%%%%
%%%%%%%%%%%%%%%%%%%%%%%%%%%%%%%%%%
\section{Introduction}\label{sec:introduction}
Are neurological networks topologically stable across
different populations of subjects or across different cognitive and
behavioral tasks? This general research program has been carried out
by a myriad of researchers in the last decade. Neuroscientists are 
often interested in evaluating whether the small-world properties of a
given brain network are conserved when comparing patients with
controls. \citet{Bassett2008}, for instance, have
studied the differences in anatomical brain networks exhibited by healthy
individuals and patients with schizophrenia. Similarly, some authors
have tested how the topological properties of certain functional networks
are affected by different behavioral tasks \citep{Heuvel2009,Fallani2008,
Cecchi2007}. Brain network topology has been studied at different
spatial scale \citep{Bassett2006}, and different time scales
\citep{Pachou2008,Salvador2008}. It is therefore undeniable that there
is considerable academic interest in comparing families of
networks; whether these represent several groups of subjects, 
or the different conditions of an experiment. This general research
paradigm is particular amenable to the analysis of subject-specific
networks. When such individual networks are available, one can readily
compute subject-specific topological measures, which will then be
compared across experimental conditions. This type of analysis has been
conducted using both functional and structural MRI data
\citep{Hagmann2008,Gong2009}. In this paper, we will mostly focus on
networks arising from functional MRI (fMRI) data.

The prospect of performing rigorous statistical 
analysis of several populations of networks, however, has been
hindered by various methodological issues. These statistical questions
have not been hitherto satisfactorily resolved in the neuroscience
community, and the field of network data analysis remains an area of
methodological development. When one is
considering the question of comparing several populations of brain 
networks, two main problems arise. First and foremost, the problem 
of the inherent dependence between connectivity strength (i.e.~ wiring density)
and network topology (i.e.~patterns of edges) necessarily arises. Most,
if not all, of the topological metrics that have become popular in the
neuroscience literature are highly sensitive to the differences in the
number of edges of the graphs under comparison. Therefore, when trying
to evaluate the topological properties of different groups of
networks on the sole basis of their topology, one also requires to
apply some level of control on the differences in density 
between the groups of networks under scrutiny. 

Secondly, the issue of separating differences in density from
differences in topology is compounded by the problem of thresholding
association matrices. In many cases, neuroscientists are considering
correlation matrices with values ranging between $-1$ and $1$. Because
network science is founded on graph theory, which is a branch of
discrete mathematics, it follows that the application of graph-theoretical
methods requires the use of a particular threshold in order to produce
adjacency matrices. Naturally, this choice of threshold is often
arbitrary, although various statistical strategies have been deployed
to alleviate the consequences of such decisions. Several authors have
thresholded correlation matrices by applying an inferential cut-off
point. This approach is similar in spirit to the standard mass
univariate strategy regularly adopted within the context of classical
statistical parametric mapping \citep{Friston1994}. 

However, this thresholding of matrices is generally critized for
throwing away valuable
information. Indeed, since network analysis proceeds by comparing
the global topological properties of the graphs obtained after
binarizing correlation matrices, it is natural to conclude that a
substantial amount of real-valued information has been discarded; and
replaced by a sequence of binary digits. As a result, several authors
have proposed to use the weighted versions of the classical
graph-theoretical measures of topology \citep{Rubinov2010}. It is
commonly believed that the use of such weighted topological statistics
alleviates both the problem of selecting an arbitrary threshold, and
also ensures that one is separating differences in topology from
differences in network density. Although this first requirement is
indeed satisfied, the second is only illusory. We will show in this
paper that the use of weighted topological measures is just as liable to be
determined by differences in density, as their standard unweighted
versions. 

%%%% 
% Paper Summary. 
In the present paper, we will concentrate our attention on weighted networks since
these are more likely to be found in the biomedical sciences than their
unweighted counterparts. This article is structured in two parts. We
firstly review how to construct summary networks representing subject-specific or
group-specific functional connectivity over time. Here, a
mass-univariate approach is adopted using different corrections for
multiple comparisons. A similar approach can also be used for representing
group differences in functional network topologies. In a second part,
we concentrate on network properties inference. This is rendered
particularly arduous by the fact that such networks tend to display
different number of edges. Since network density is highly predictive
of a host of network topological measures, such statistical inference requires special
attention, when comparing groups of subjects that exhibit substantial
differences in network density. 

%%%%%%%%%%%%%%%%%%%%%%%%%%%%%%%%%%
%%%%%%%%%%%%%%%%%%%%%%%%%%%%%%%%%%
%%%%%%%%%%%%%%%%%%%%%%%%%%%%%%%%%%
\section{Construction of Summary Networks}\label{sec:construction}
We firstly describe how one can construct summary networks from a
family of subject-specific weighted or unweighted networks. This task
can be tackled by combining the data available, using a
mass-univariate approach, as is commonly done in fMRI. Note that the
phrases, graph and network, will be used interchangeably in this
paper. 

%%%%%%%%%%%%%%%%%%%%%%%%%%%%%%%%%%
\subsection{Statistical Parameter Networks (SPNs)}\label{sec:spn}
Here, we review an efficient method for summarizing inference on networks, using a
mass-univariate approach. By tacit consensus, this method has
essentially become the norm in the field
\citep{Achard2006,Ginestet2012a,He2007,He2009}.
This strategy should be compared to the 
one adopted in the classical statistical parametric mapping (SPM)
framework, which has been utilized in neuroimaging for the past two decades
\citep{Friston1994}. Consequently, this approach will be 
referred to as statistical parametric networks
(SPNs). The problem of constructing a summary graph centers on how to combine
the elements of a population of subject-specific correlation matrices.
In the SPN framework, summary networks are constructed irrespective of
whether or not structural or functional data are being used. 
While in fMRI studies, it has been common for researchers to 
compute correlations over time between regions of interest
\citep{Achard2006,Achard2007}, studies based on structural MRI data,
by contrast, have considered between-regions correlations with respect
to the available population of subjects \citep{Bassett2008}. 
In this section, we will concentrate on the specific problem posed by the
study of functional MRI cortical networks, where each subject-specific
correlation matrix represent inter-regional normalized covariances, computed
with respect to a sequence of time points. 

Succinctly, one may say that an SPN is to a correlation matrix, what
an SPM is to an intensity map. As for the latter, an SPN can be produced in order to 
obtain a summary network. Different summary networks can be
constructed for the different conditions of an experiment, or for the different
groups of subjects under scrutiny. \citet{Achard2006} and \citet{He2009}, for
instance, have visualized their data using summary networks, whereby
an edge is solely included when a corresponding
test statistic for that edge is significant. We will refer to such
summary networks as \ti{mean} SPNs. Similarly, one can construct
\ti{differential} or \ti{difference} SPNs, which represent the edges
that have been significantly `lost' and the edges that have been
significantly `gained', when comparing the graphs across
experimental conditions, or when considering several groups of
subjects. Under its many guises, this approach has been adopted by
various authors including \citet{Zalesky2010} and
\citet{Richiardi2011}, who have used network-based statistics and
machine learning methods, respectively, for the comparison of a group
of subjects with a group of controls. 

The SPN approach that we wish to present here is slightly more general, since it 
accommodates sophisticated experimental designs, in which information may be pooled over
a number of experimental conditions. As for SPM, such analyses  
enable a concise visualization of the data, which can be interpreted
in terms of network properties, topology and community structure. 
This approach is particularly helpful for an efficient reporting
of the experimental results. As mentioned in the introduction, the use
of SPNs has the additional advantage of somewhat alleviating the
methodological concerns associated with the choice of an arbitrary
threshold value; since we are here selecting such cut-off points on the
basis of a specific $p$-value. Network thresholding is
therefore here supplanted by inference.

\begin{figure}
  \footnotesize \centering
  \tb{(A)} \\
  %vspace{.25cm}
  $0$-back \hspace{1.0cm} $1$-back \hspace{1.0cm} $2$-back
  \hspace{1.0cm} $3$-back \\
  \vspace{.1cm}
  \includegraphics[width=2.0cm]{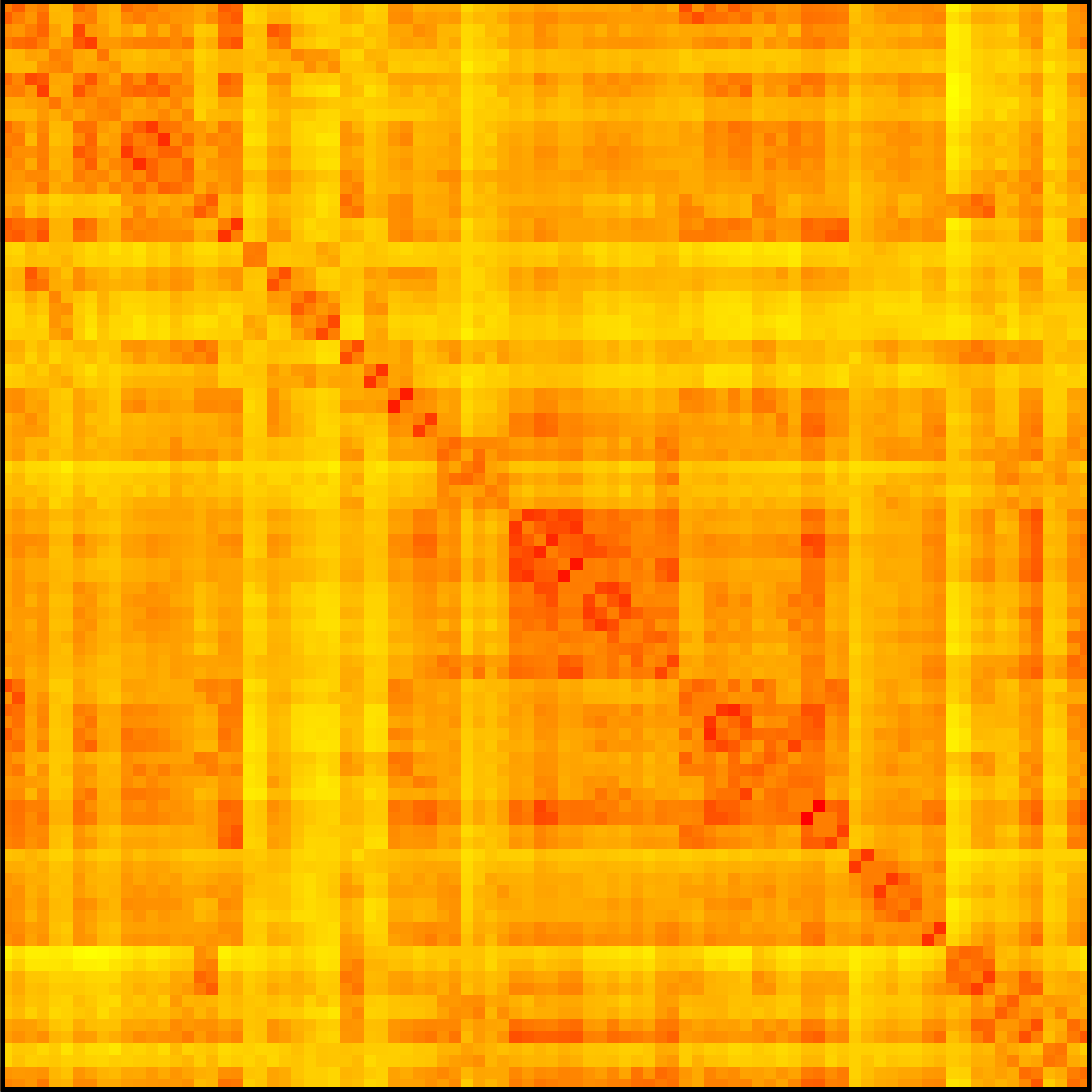}
  \includegraphics[width=2.0cm]{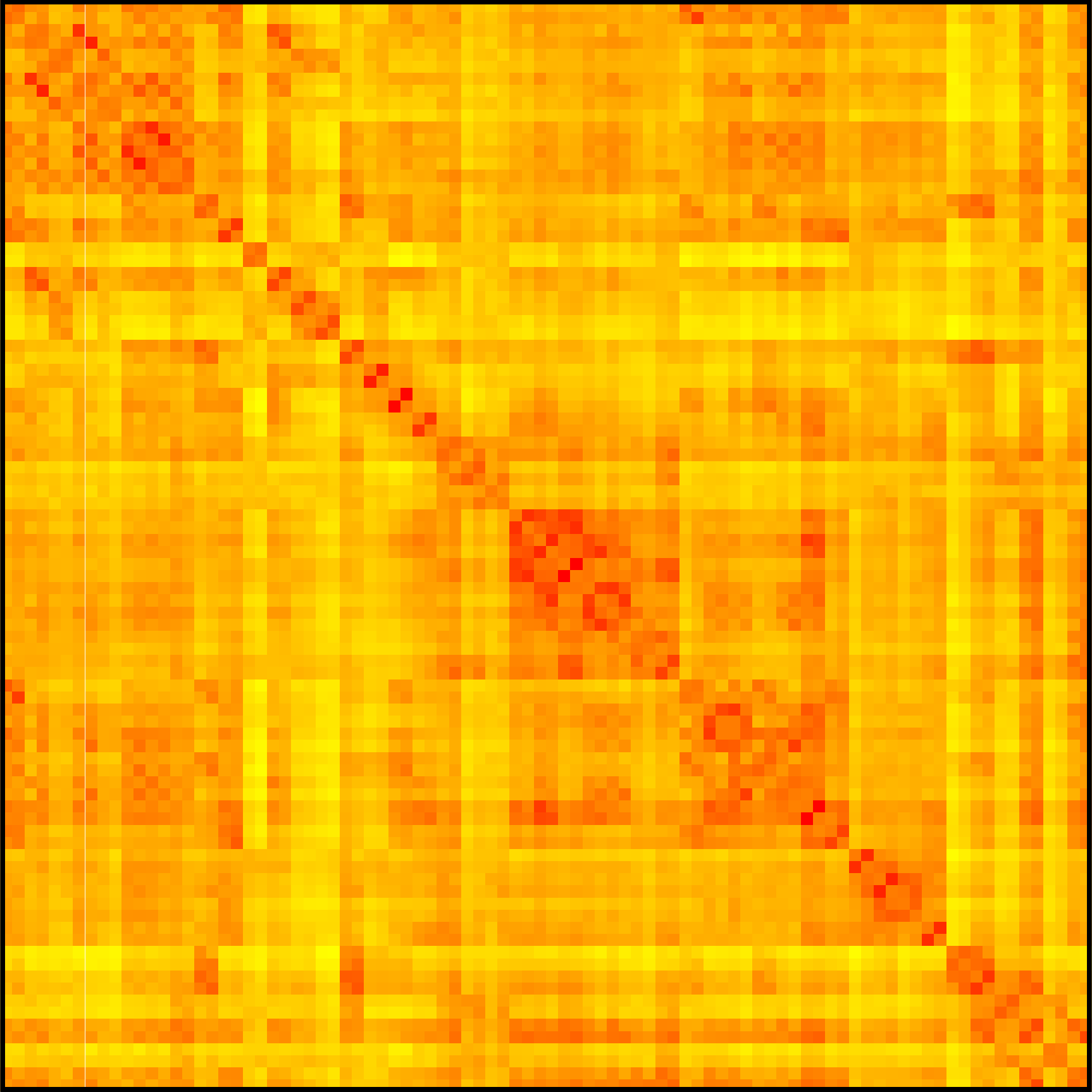}
  \includegraphics[width=2.0cm]{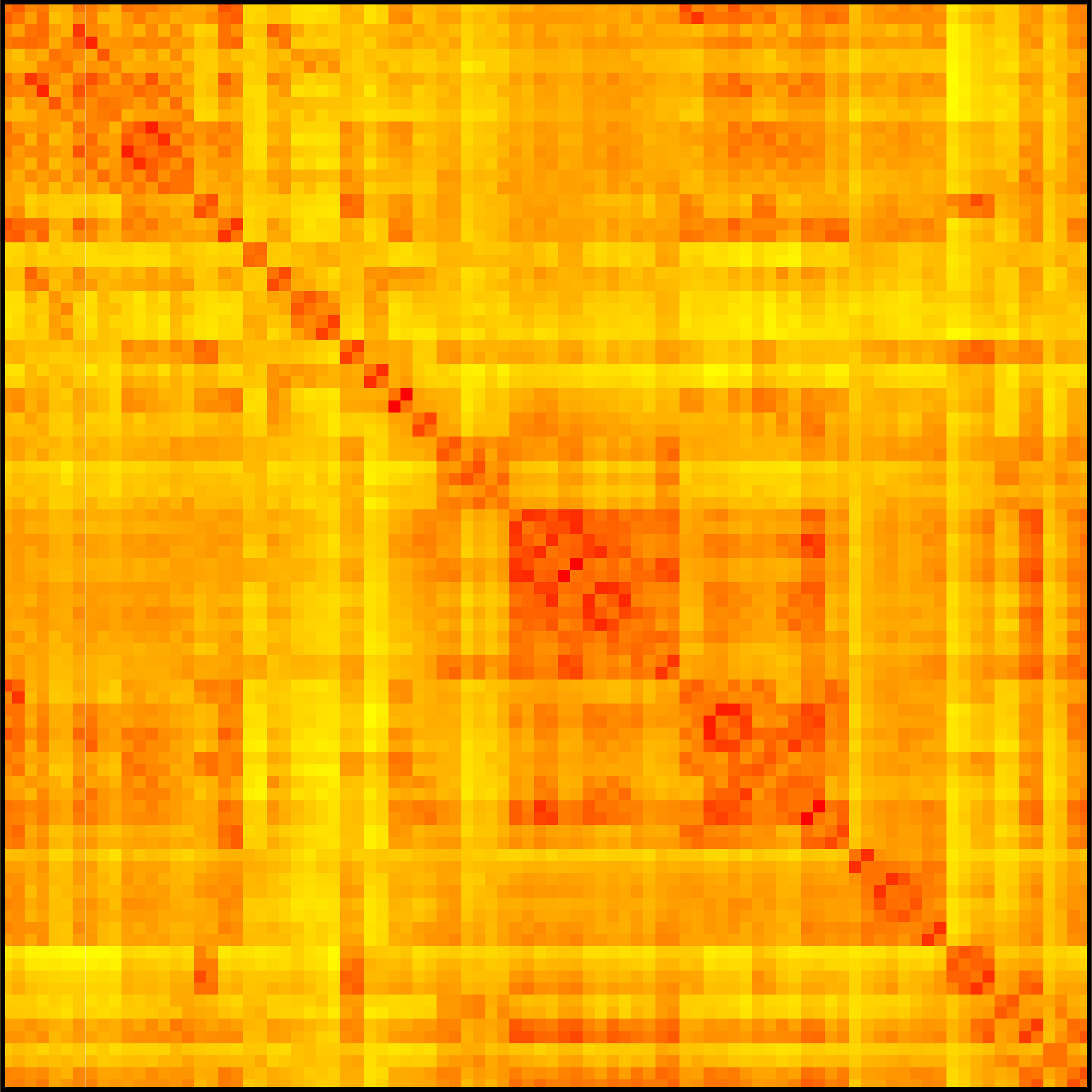}
  \includegraphics[width=2.0cm]{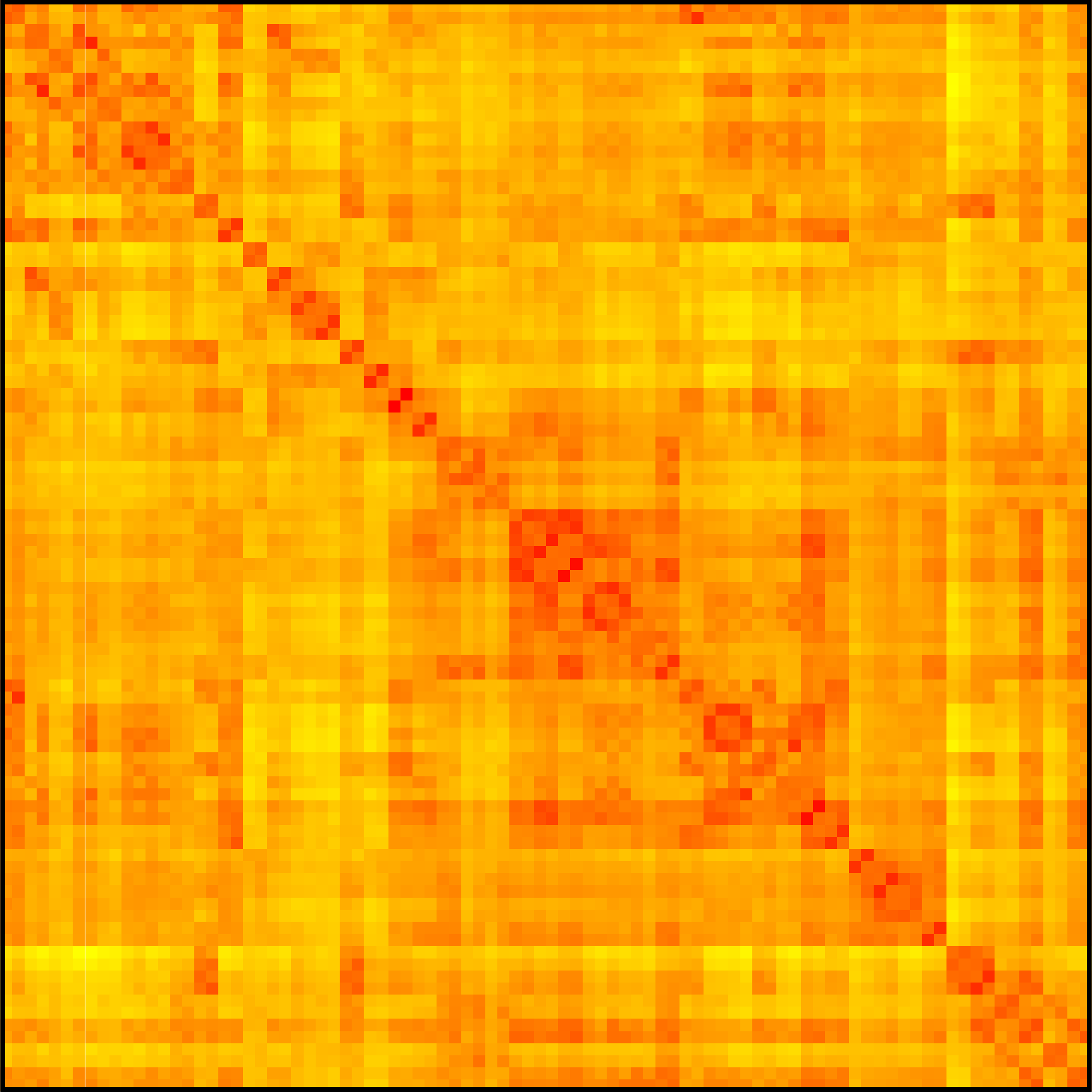}\\
  \includegraphics[width=4.5cm]{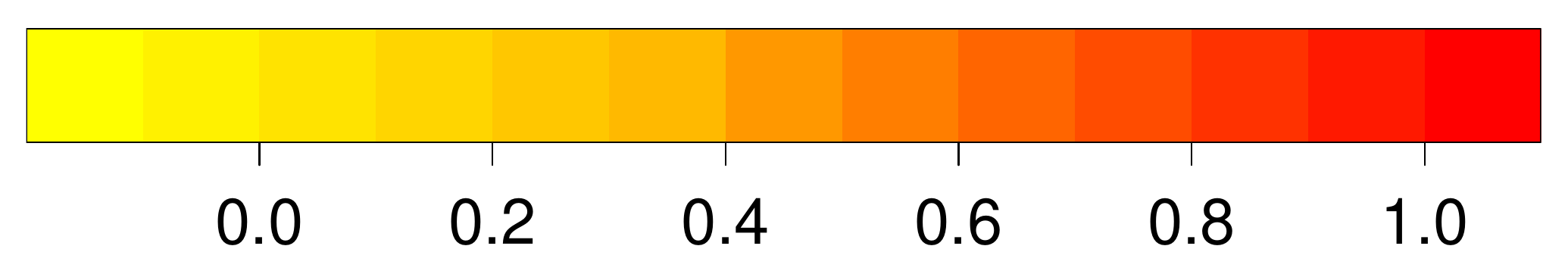}\\
  \vspace{.25cm}
  \tb{(B)} \\
  %vspace{.25cm}
  $0$-back \hspace{1.0cm} $1$-back \hspace{1.0cm} $2$-back
  \hspace{1.0cm} $3$-back \\
  \vspace{.1cm}
  \includegraphics[width=2.0cm]{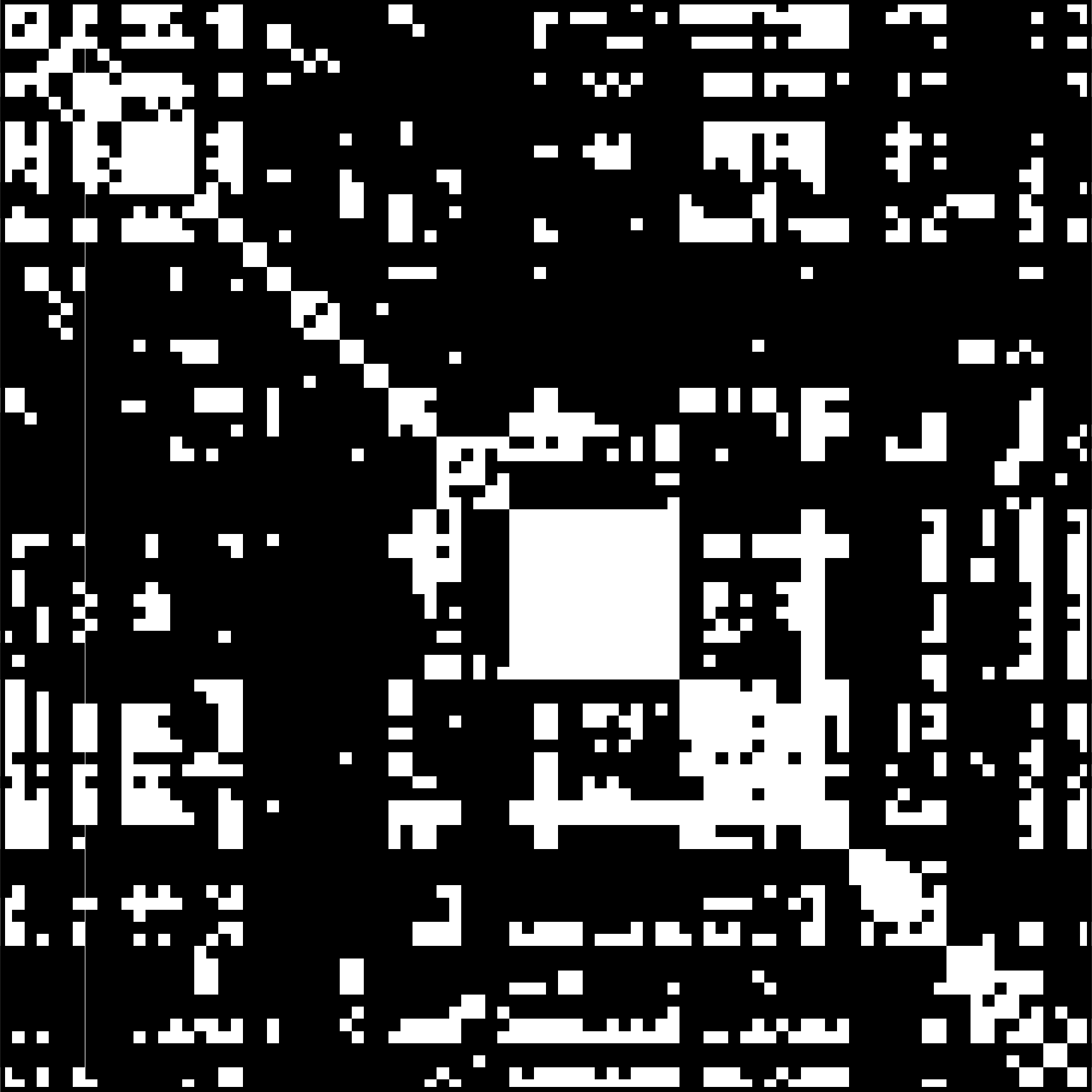}
  \includegraphics[width=2.0cm]{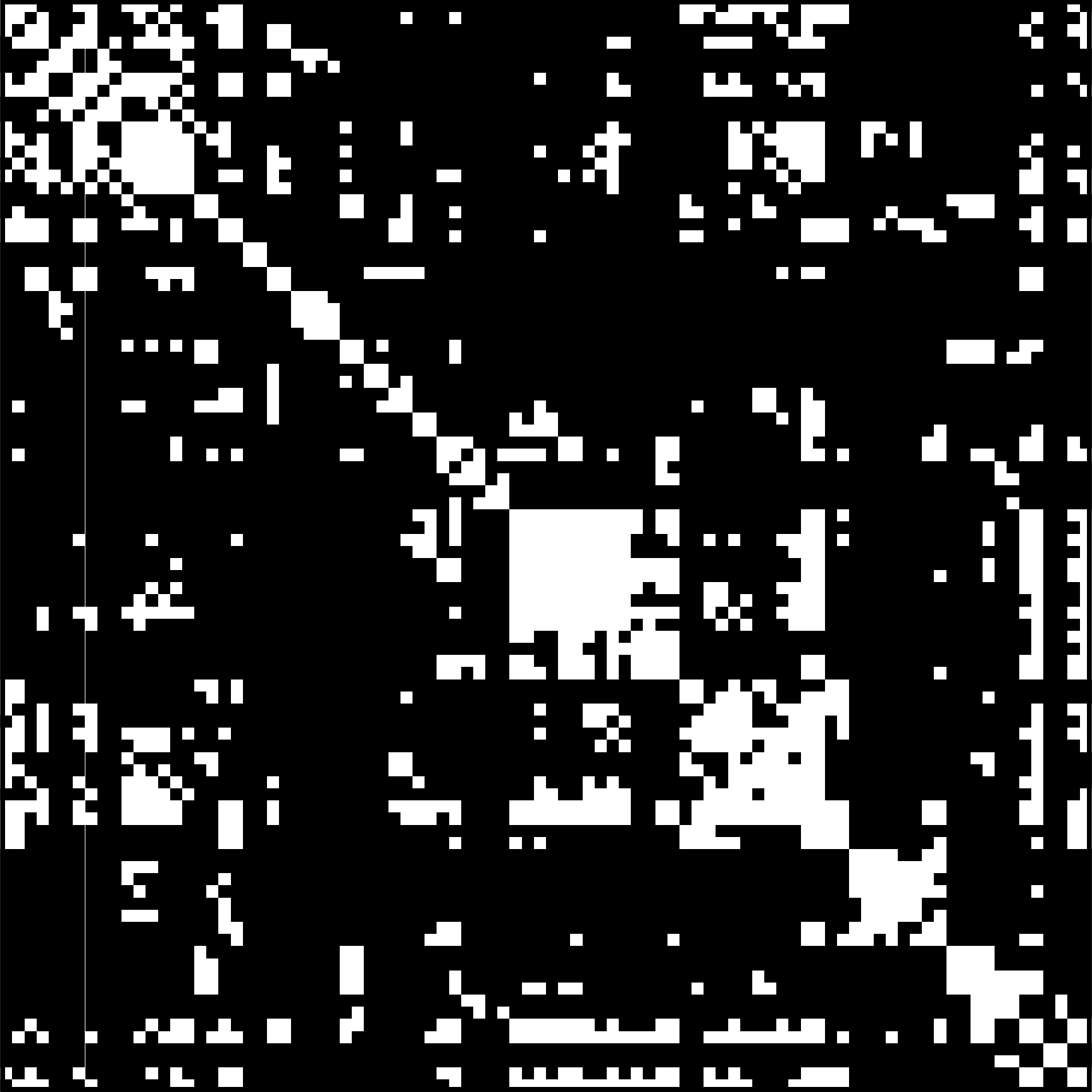}
  \includegraphics[width=2.0cm]{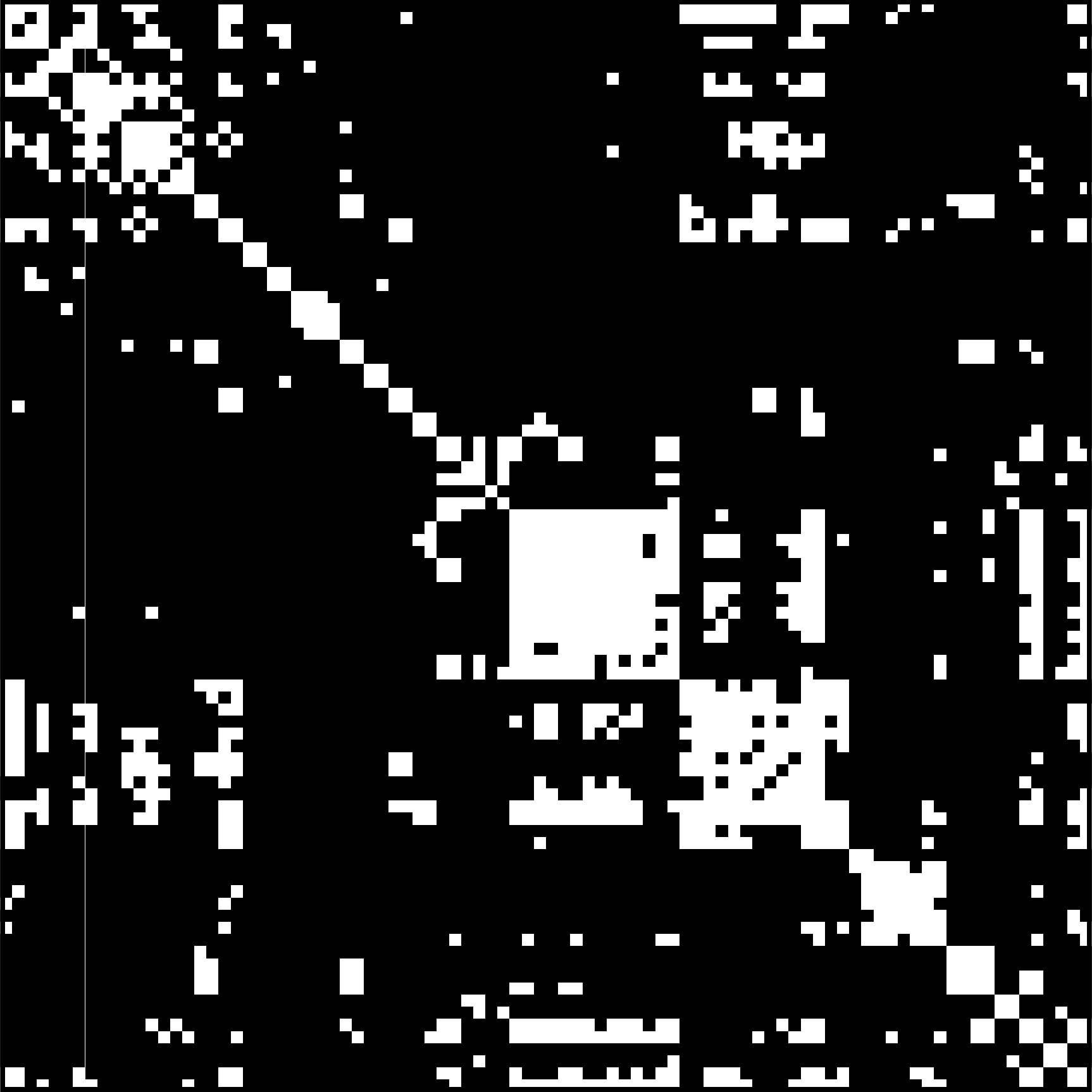}
  \includegraphics[width=2.0cm]{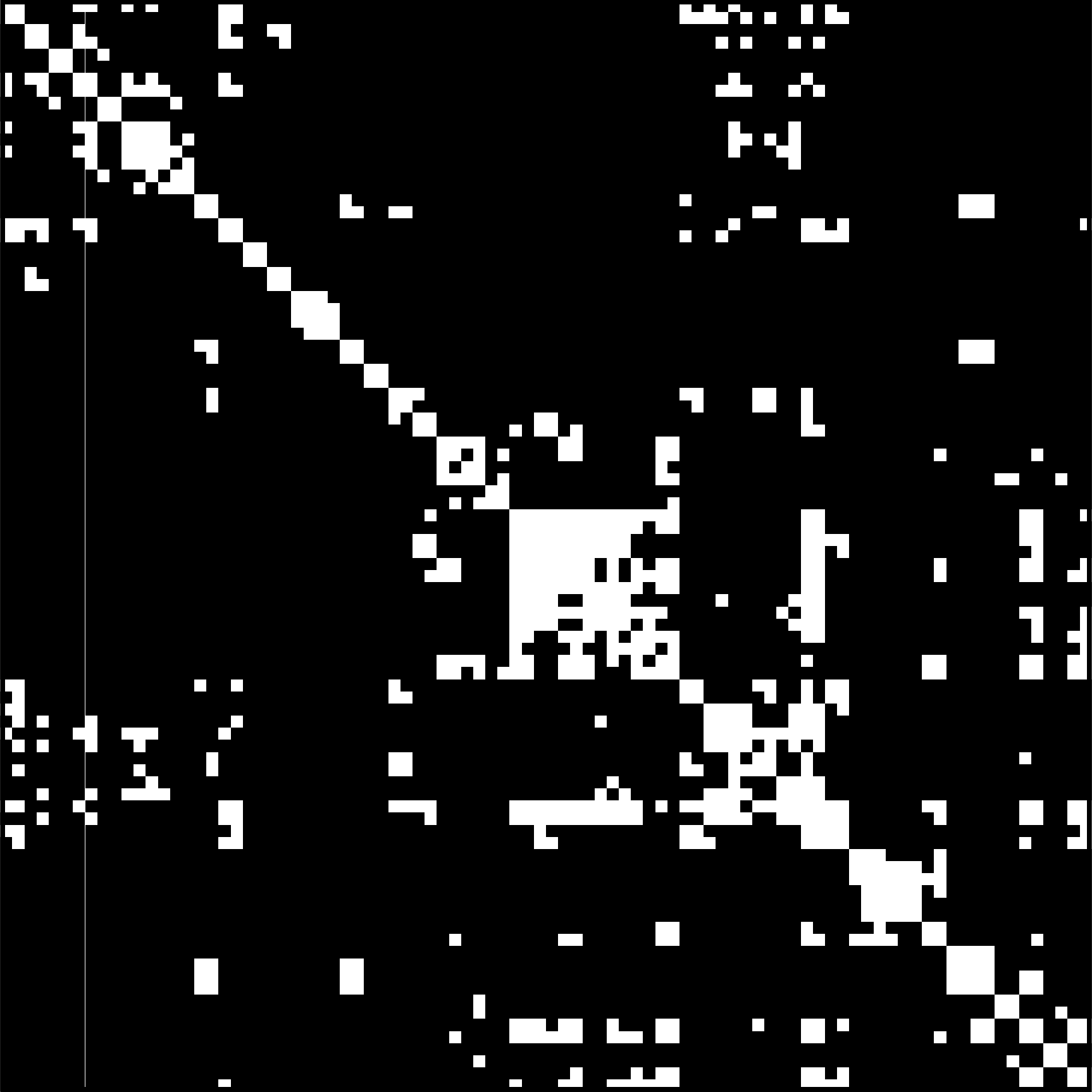}\\
  \caption{\textbf{Illustration of the use of mean SPNs to summarize
      networks in a cognitive task.} Adjacency matrices become sparser with
    increasing working memory load.
    In panel (a), heatmaps corresponding to
    the correlation matrices in each of four $N$-back conditions, for $n=43$
    subjects. In panel (b), the adjacency
    matrices were obtained by constructing mean SPNs, 
    using a mass-univariate approach based on $z$-tests with respect
    to the grand sample mean $\bar{\bar{r}}$ and the grand sample
    standard deviation  $\op{sd}(\br)$ with 
    FDR correction (base rate $\alpha_{0}=.05$). 
    Zero entries are denoted in black in the adjacency
    matrices. (See \citet{Ginestet2011a} for a full description.)
    \label{fig:heatmap}}
\end{figure}

\sub{Statistical Parametric Networks (SPNs)}
The thresholding of association matrices, such as correlation
matrices, is equivalent to the application of an elementwise indicator
function. This type of function, however, is non-linear, in the sense
that the sum of the thresholded correlation matrices is not equal to
the thresholded mean correlation matrix. That is, this may be formally
expressed, as follows,
\begin{equation}
     \sum_{i=1}^{n}T_{\tau}(\bR_{i}) \neq T_{\tau}\lt(
     \sum_{i=1}^{n}\bR_{i}\rt),
     \label{eq:quasilinearity}
\end{equation}
where $i=1,\ldots,n$ labels the subjects taking part in the
experiment, and where $\bR_{i}$'s denote subject-specific correlation
matrices. Here, the function, $T_{\tau}$, is a thresholding function that
takes a matrix, and returns its binarized version, with respect to a
cut-off point, $\tau$. The issue of thresholding correlation matrices is
illustrated in figure \ref{fig:heatmap}, where we have reported some
of the preliminary data analysis conducted in \citet{Ginestet2012a}. 

Currently, there is little guidance on how one should proceed, when 
summarizing the network analysis of a given study. There is hence a 
pressing need to reach a methodological consensus on how to
standardize the construction and reporting of summary networks in
neuroscience. A natural desideratum for such summary networks is that
they should reflect the topological variability of the entire 
population of networks. Pioneering work in that direction has been
laid out by several authors, including \citet{Achard2006} and
\citet{He2009}, for the consideration of a single family of graphs.
In the sequel, we review these ideas and extend them 
to the case of several populations of networks, as was conducted in 
\citet{Ginestet2012a}. 
%%%%%%%%%%%%%%%%%%%%%%%%%%%%%%%%%%%%%%%%%%%%%%%%
% HeatmapBs. 
%%%%%%%%%%%%%%%%%%%%%%%%%%%%%%%%%%%%%%%%%%%%%%%%

The question of drawing inference on families of networks that vary
over several experimental conditions can be subdivided into two
related issues. On the one hand, one needs to test whether or not the properties
of the nodes have been significantly affected by the experimental manipulation. On
the other hand, one also needs to evaluate whether or not the presence
and absence of edges have significantly varied across the experimental
conditions. One can drawn statistical inference for these two
distinct, yet related, research questions. Contrary to the classical
SPM framework, these two distinct problematics need to be 
answered using two different types of networks: one for comparing
vertices, and another for comparing edges. 

A substantial advantage of the SPN methodology is that it addresses the
problem arising from the quasi-linearity of the thresholding function
presented in equation (\ref{eq:quasilinearity}). Indeed, since we are
drawing inference using the correlation coefficients per se, we
consequently bypass the problem of averaging over a set
of thresholded correlation matrices; while nonetheless producing
a statistical summary taking the form of a graph.

We here employ standard graph theoretical notation in order formulate
our approach to this specific problem. The interested reader is
invited to consult \citet{Bollobas1998} for a more solid introduction
to graph objects and their properties. As aforementioned, we will here
use the terms networks and graphs interchangeably. In the context of discrete
mathematics, a graph $G$ is formally defined as an ordered pair of
sets $(V,E)$; in which $V(G)$ represents the set of \ti{vertices}
(sometimes referred to as nodes) in the graph of interest; whereas
$E(G)$ denotes the set of \ti{edges} in that network (also called connections). 
The total number of edges and total number of nodes in $G$ will be
concisely denoted by $N_{E}$ and $N_{V}$, respectively. A one-way
experimental design may be typically composed of $J$ experimental
conditions, with $n$ subjects, per experiment. Thus,
the full data set of interest can be described as an $(n\times J)$-matrix
of correlation matrices. In the sequel, the indexes $i=1,\ldots,n$
will label the experimental subjects; whereas the indexes
$j=1,\ldots,J$ will refer to the experimental conditions.
Formally, one could represent the full data set as the
following matrix, 
\begin{equation}
  \bR = 
  \begin{vmatrix}
       \bR_{11} & \hdots    &\bR_{1J} \\
       \vdots   & \ddots   & \vdots  \\
       \bR_{n1} & \hdots  & \bR_{nJ} \\
  \end{vmatrix}.
  \label{eq:bigR}
\end{equation}
Here, each element $\bR_{ij}$ in this equation denotes 
a correlation matrix of dimension $N_{V}\times N_{V}$. There is a
one-to-one correspondence between each of these correlation matrices
and a weighted graphs on $N_{V}$ vertices or nodes. The individual
vertices will be labeled by $v=1,\ldots,N_{V}$. Moreover, 
for convenience, each of the matrix entries in $\bR$, will be denoted by
$r^{e}_{ij}$; where the superscript $e$ labels an edge from the
saturated or complete graph, which possesses the maximal number of possible
edges. That is, the saturated graph has the following edge set size,
$N_{V}(N_{V}-1)/2$. In the rest of this paper, edges will
be systematically referred to by using superscripts.

\begin{figure}[t]
  \footnotesize
  % \small
    \centering
    %%%%%%%%%%%%%%%%%%%%%%%%%%%%%%%%%%%%%%%%%%%%%%%%%%
    \tb{(a) Coronal ${\op{SPN}}_{j}$}\\
    \vspace{.1cm}
    \ti{Sup.}\hspace{2.5cm}
    \ti{Sup.}\hspace{2.5cm}
    \ti{Sup.}\hspace{2.5cm}
    \ti{Sup.}\\
    \includegraphics[width=2.7cm]{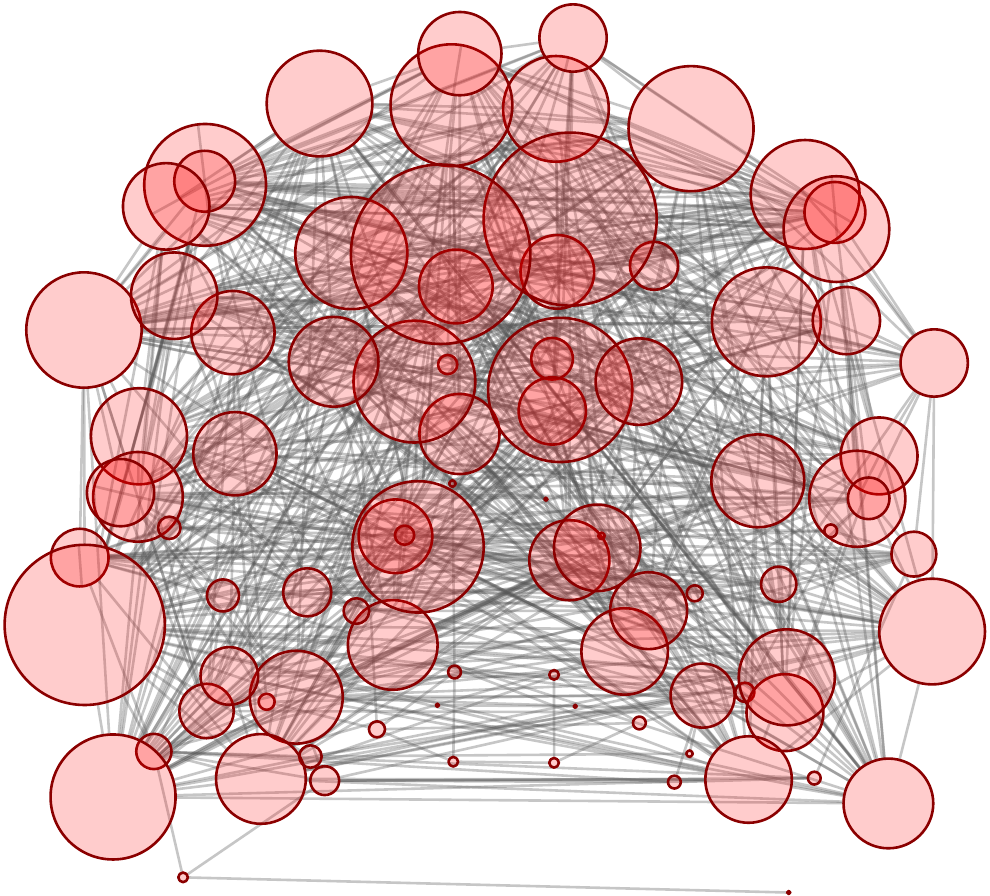}\hspace{.3cm}
    \includegraphics[width=2.7cm]{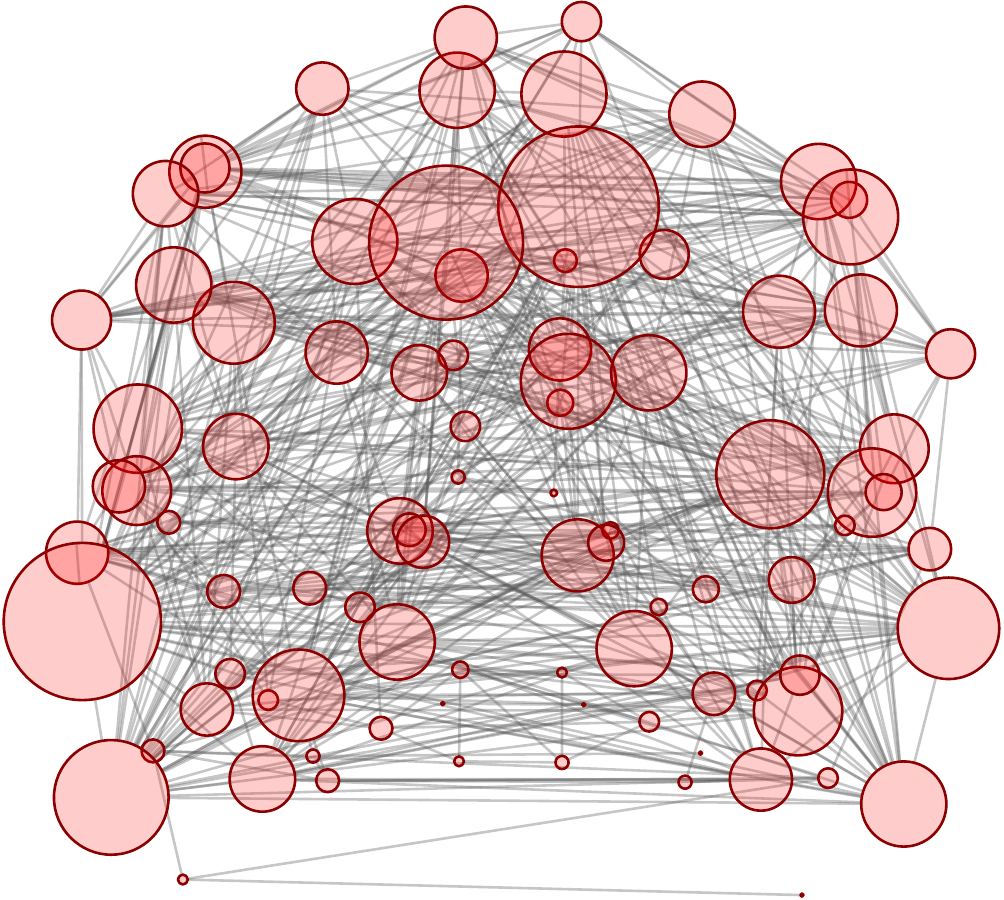}\hspace{.3cm}
    \includegraphics[width=2.7cm]{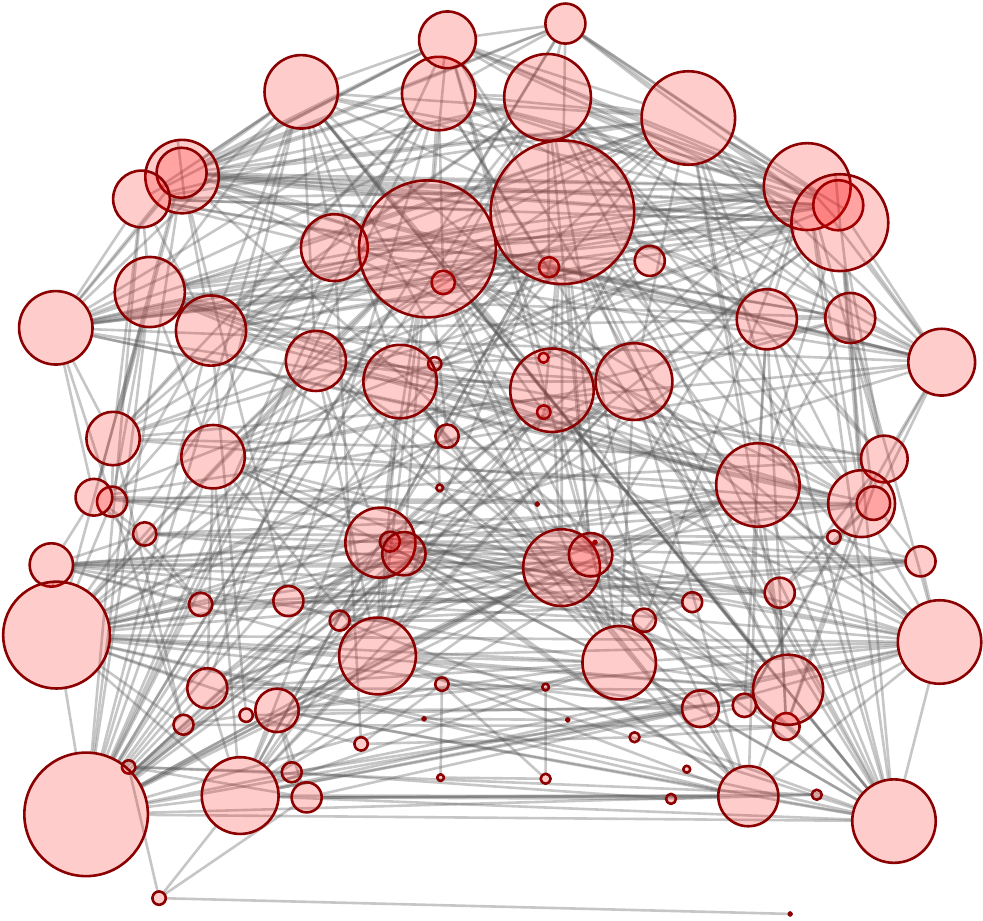}\hspace{.3cm}
    \includegraphics[width=2.7cm]{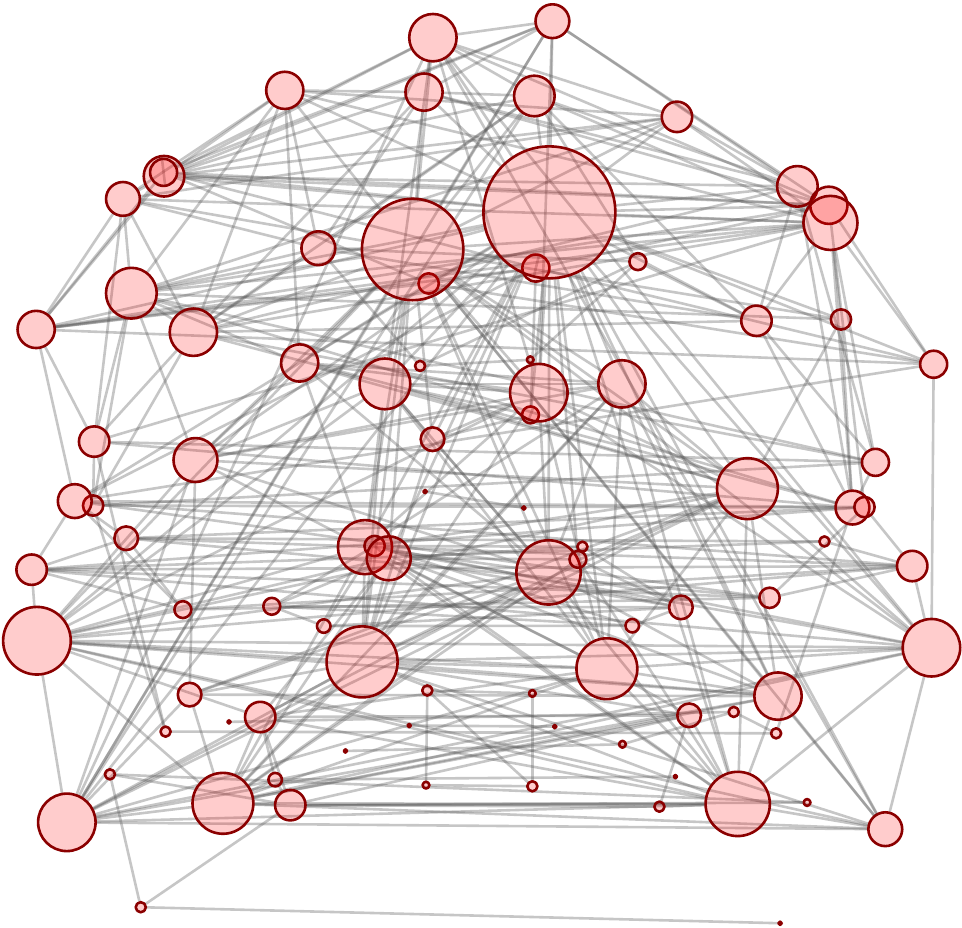}\\
    \ti{Inf.}\hspace{2.5cm}
    \ti{Inf.}\hspace{2.5cm}
    \ti{Inf.}\hspace{2.5cm}
    \ti{Inf.}\\
    %%%%%%%%%%%%%%%%%%%%%%%%%%%%%%%%%%%%%%%%%%%%%%%%%%
    \vspace{.5cm}
    \tb{(b) Transverse ${\op{SPN}}_{j}$}\\
    \vspace{.1cm}
    \ti{Ant.}\hspace{2.5cm}
    \ti{Ant.}\hspace{2.5cm}
    \ti{Ant.}\hspace{2.5cm}
    \ti{Ant.}\\
    \includegraphics[width=2.7cm]{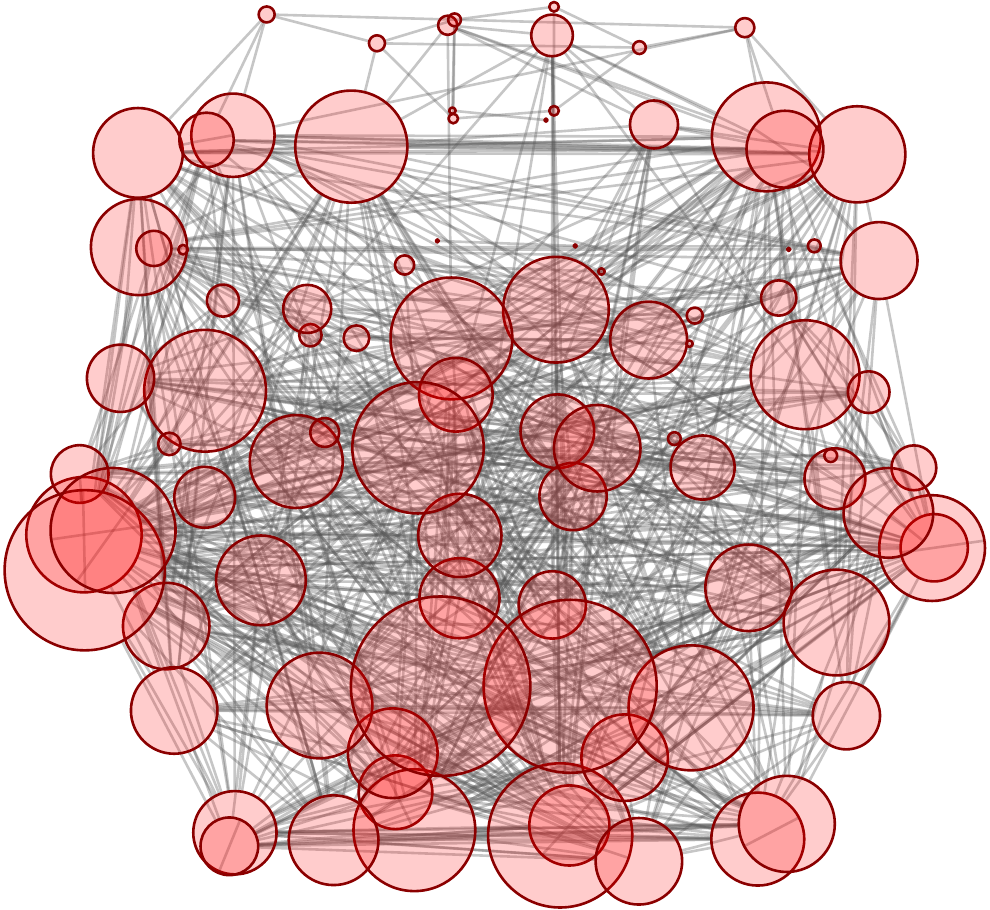}\hspace{.3cm}
    \includegraphics[width=2.7cm]{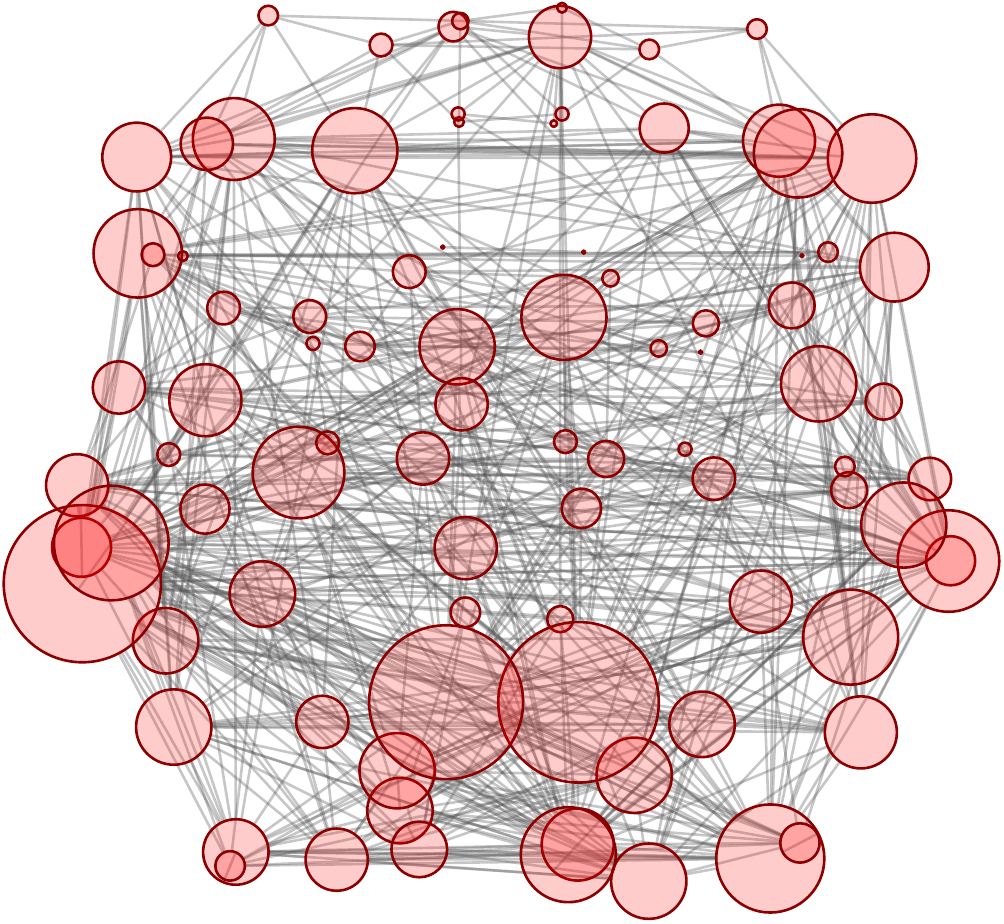}\hspace{.3cm}
    \includegraphics[width=2.7cm]{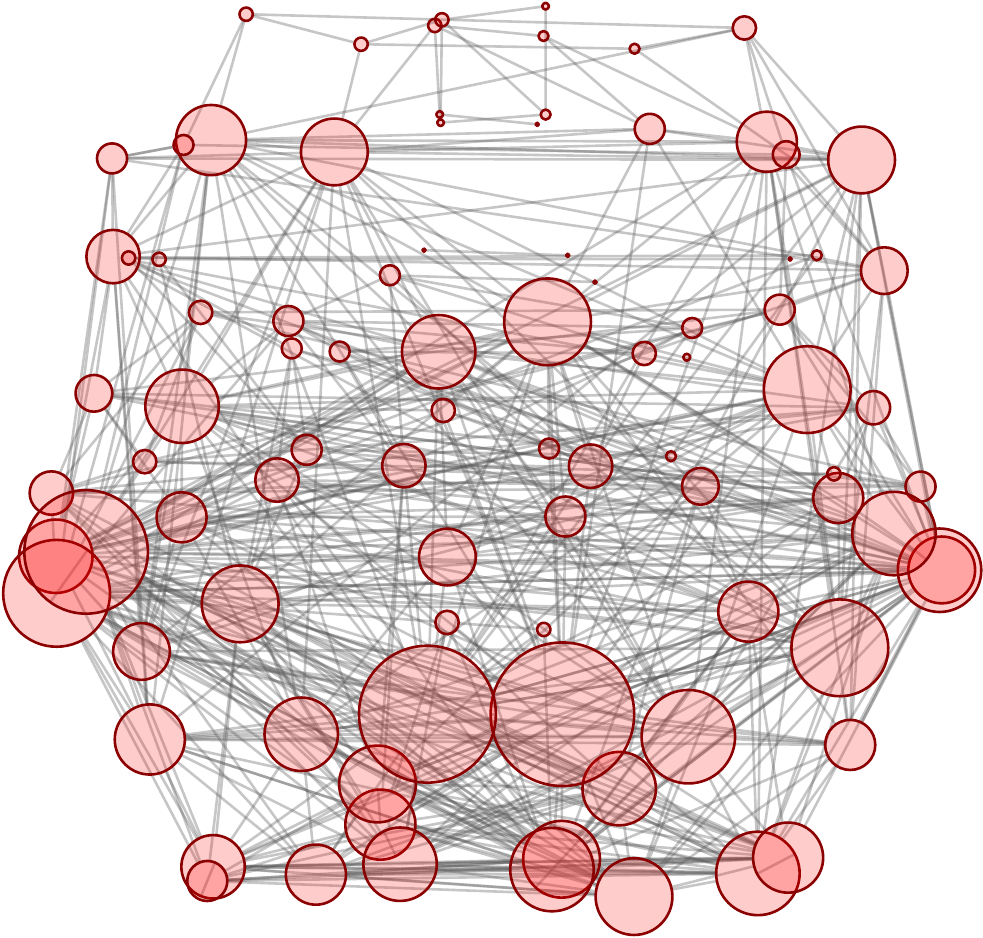}\hspace{.3cm}
    \includegraphics[width=2.7cm]{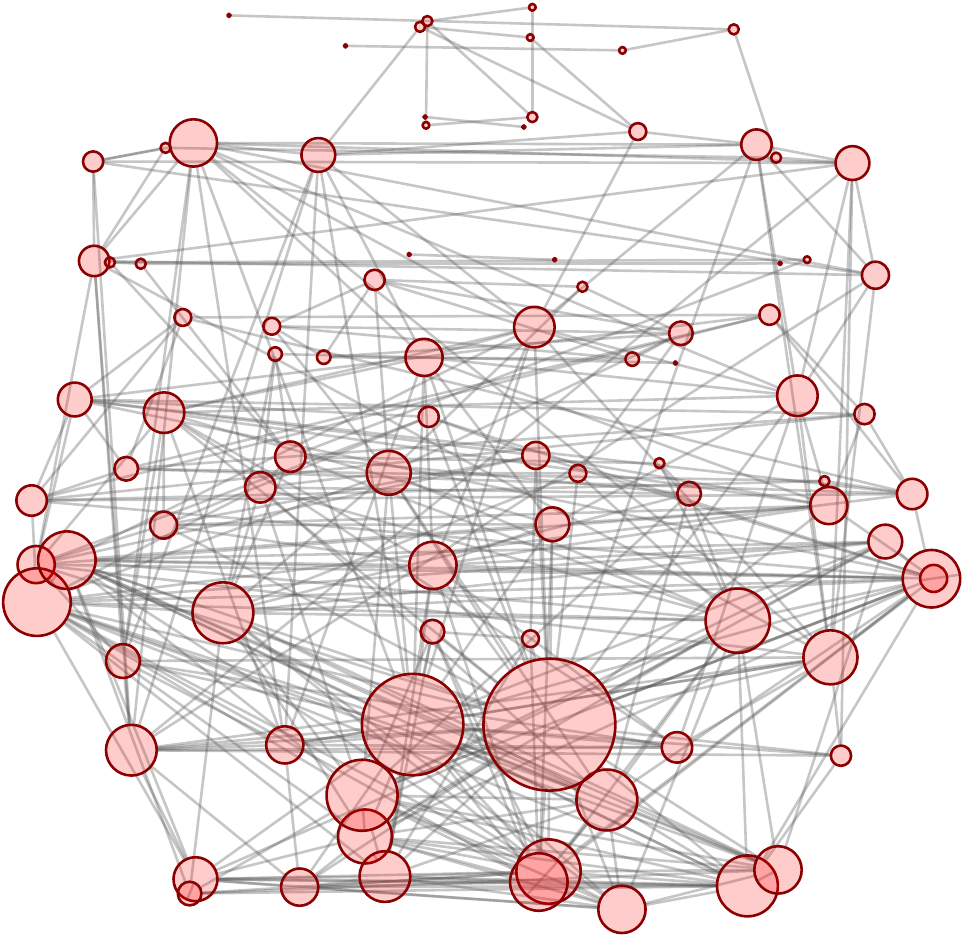}\\
    \ti{Post.}\hspace{2.5cm}
    \ti{Post.}\hspace{2.5cm}
    \ti{Post.}\hspace{2.5cm}
    \ti{Post.}\\
    %%%%%%%%%%%%%%%%%%%%%%%%%%%%%%%%%%%%%%%%%%%%%%%%%%
    \vspace{.5cm}    
    \tb{$0$-back}\hspace{2.3cm}
    \tb{$1$-back}\hspace{2.3cm}
    \tb{$2$-back}\hspace{2.3cm}
    \tb{$3$-back}
    \caption{\textbf{Graphical representations of mean SPNs over four
        levels of a cognitive task.} The mean SPNs for an $N$-back
     task, in the coronal (a) and transverse (b) planes are here presented,
     after FDR correction (base rate $\alpha_{0}=.05$). Locations of the nodes correspond to
      the stereotaxic centroids of the corresponding cortical regions. The
      orientation axes are indicated in italics: inferior--superior
      and anterior--posterior for the coronal and transverse sections,
      respectively. The size of each node is proportional to its
      degree. (See \citet{Ginestet2011a} for a full description.)
      \label{fig:net_array}}
\end{figure}

%%%%%%%%%%%%%%%%%%%
\sub{Mean SPNs}
A mean or summary SPN allows to statistically infer the
`average' set of inter-regional connections in a group of subjects. 
Such SPNs are generally obtained by adopting a mass-univariate
approach, whereby a sequence of statistical tests are performed for
each edge in the edge set. Such an operation may be repeated for each
experimental condition. Using the notation introduced earlier, one may
conduct a test for each of the columns in the array, denoted $\bR$, in equation
(\ref{eq:bigR}). In effect, we are here considering the following column vectors
of correlation matrices, 
\begin{equation}
    \bR_{j} = [\bR_{1j},\ldots,\bR_{nj}]^{T}.
\end{equation}
Each of these column vectors is analyzed independently in order to
produce a single network for each of the different experimental
conditions. For the case of correlation matrices, the original matrix
entries are routinely Fisher $z$-transformed, in order to be able to
use central limit theorems for approximating the density functions of these
test statistics. In doing so, one can then draw inference, using 
an analysis of variance, for instance, or another adequate statistical
model, suitable for the data at hand. An example of such mean SPNs under
different experimental conditions is reported in figure
\ref{fig:net_array}. 

%%%%%%%%%%%%%%%%%%%%%%%%%%%%%%%%%%%%%%%%%%%%%%%%
% Set of Nets.
%%%%%%%%%%%%%%%%%%%%%%%

%%%%%%%%%%%%%%%%%%%
\sub{Differential SPNs}
Perhaps, the tenor research question in network data analysis in
neuroscience is whether certain edges have been `gained' or `lost', 
as a consequence of a particular experimental condition. 
This general research question can be specifically answered by
computing two distinct differential networks, representing what we may
call the \textit{downweighted} and \textit{upweighted} SPNs. These two
types of networks will be denoted by $\op{SPN}_{-}$ and
$\op{SPN}_{+}$, respectively. 

As for mean SPNs, the construction of these differential
networks can similarly be conducted within a mass-univariate
approach. For differential SPNs, however, statistical inference needs
to be drawn from the full data set. That is, one needs to consider all
the correlation coefficients described in equation (\ref{eq:bigR})
--that is, the elements contained in the matrix $\bR$. 
Computing a differential SPN will generally involve $N_{E}$ linear
models. Depending on the general experimental framework adopted by the
researchers, these linear models could be extended to mixed effects
models. In its most general formulation, we may consider a repeated
block design, which can be succinctly expressed by using the classical
formalism due to \citet{Laird1982},
\begin{equation}
     \br_{i}^{e} =  \bX^{e}_{i}\bbe^{e} + \bZ^{e}_{i}\Bb_{i}^{e} + \bep^{e}_{i};
     \qq i=1,\ldots,n. \label{eq:glm1}
\end{equation}
Here, each vector, $\br_{i}^{e}=[ r_{i1}^{e},\ldots,r_{iJ}^{e}]^{T}$, denotes the
correlation coefficients of interest, and $\bbe^{e}=[
\beta_{1}^{e},\ldots,\beta_{J}^{e}]^{T}$ consists of the vector of fixed
effects. The latter does not vary over subjects and will be the main
object of study. By contrast, the $\Bb^{e}_{i}$'s are the vector of 
subject-specific random effects, which will be integrated
over. Finally, $\bep^{e}_{i}=[\epsilon_{i1}^{e},\ldots,\epsilon_{iJ}^{e}]^{T}$ is the
vector of residuals. Crucially, the $\bX_{i}$'s and $\bZ_{i}$'s denote
the design matrices for the fixed and random effects, respectively.
As in standard applications of mixed effects models, the
covariance matrices for $\bep^{e}$ and $\Bb^{e}$ can be assumed to be
diagonal and positive semi-definite, respectively \citep[see][for
details]{Demidenko2004}.

In general, one may include an edge in a differential SPN,
when the corresponding $F$-test for the experimental factor has been found to be
significant. Depending on the linear model used, different
statistical test may be performed \citep{Pinheiro2000}. 
Therefore, the use of a mass-univariate approach for extracting
between-condition differences in the presence or absence of edges, 
yields two different types of differential SPNs. That is, depending on
the sign of the significant fixed effect coefficients, one may include
that edge in either a downweighted network, which may be denoted
$\op{SPN}_{-}$; or in an upweighted network, denoted
$\op{SPN}_{+}$. 

A similar approach can be adopted to estimate the upweighting and downweighting of
the signal of interest at single \ti{nodes}. Again, such a
node-specific differential SPN can be obtained by performing a set of $N_{V}$
linear models. In this case, the data under consideration is the
set of matrices $\bY^{v}=\lb y_{ij}^{v}\rb$, where each $v\in V$ is a
region of interest. Every $y_{ij}^{v}$ corresponds to a time-averaged
intensity signal, for the $v\tth$ region, for subject $i$,
under the $j\tth$ experimental condition. Thus, one could reformulate
the system of equations for evaluating edges in (\ref{eq:glm1}) by 
using superscripts to denote vertices. 

\begin{figure}[t]
   \centering
   \includegraphics[width=6.5cm]{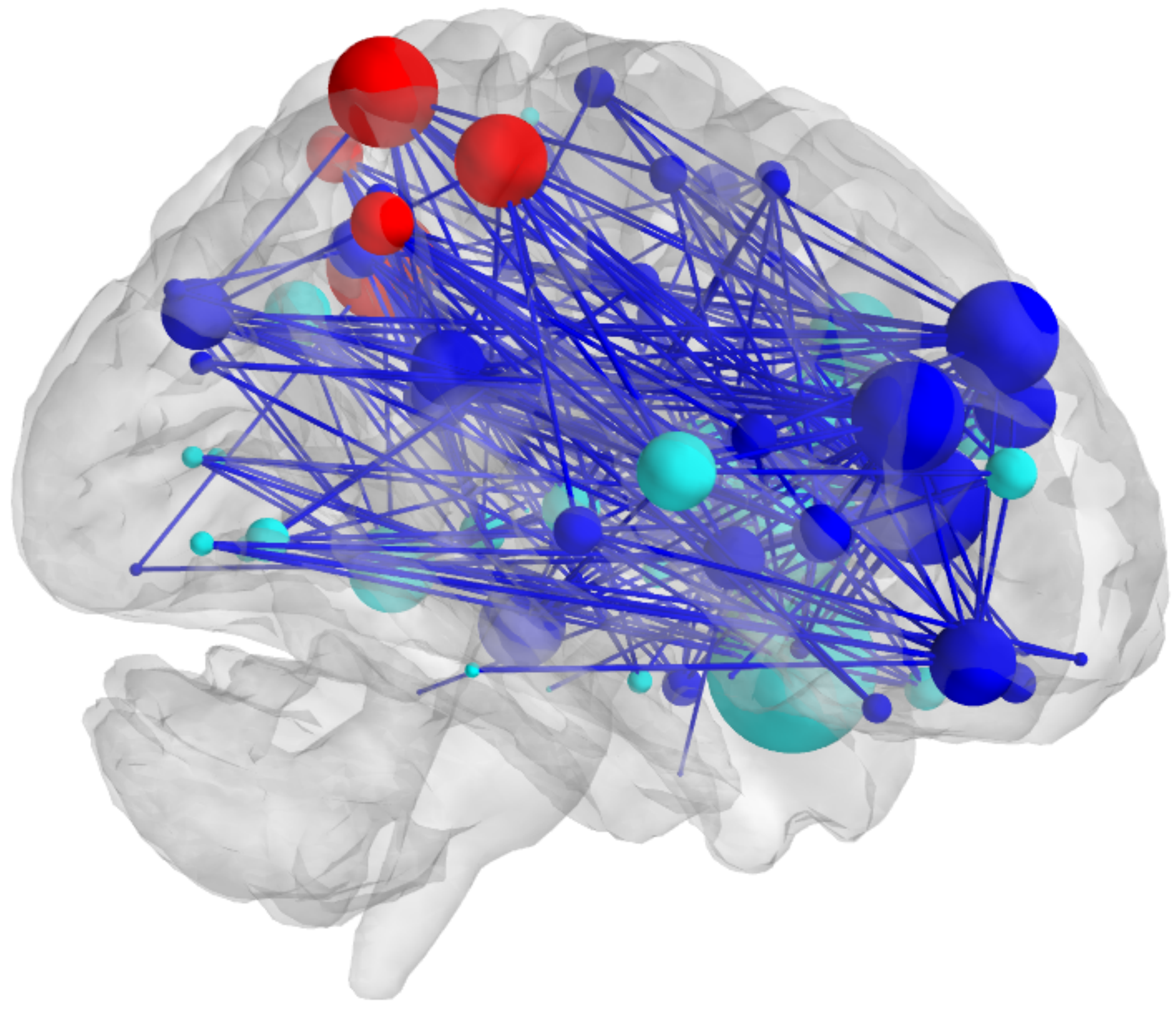}
   \hspace{.5cm}
   \includegraphics[width=6.5cm]{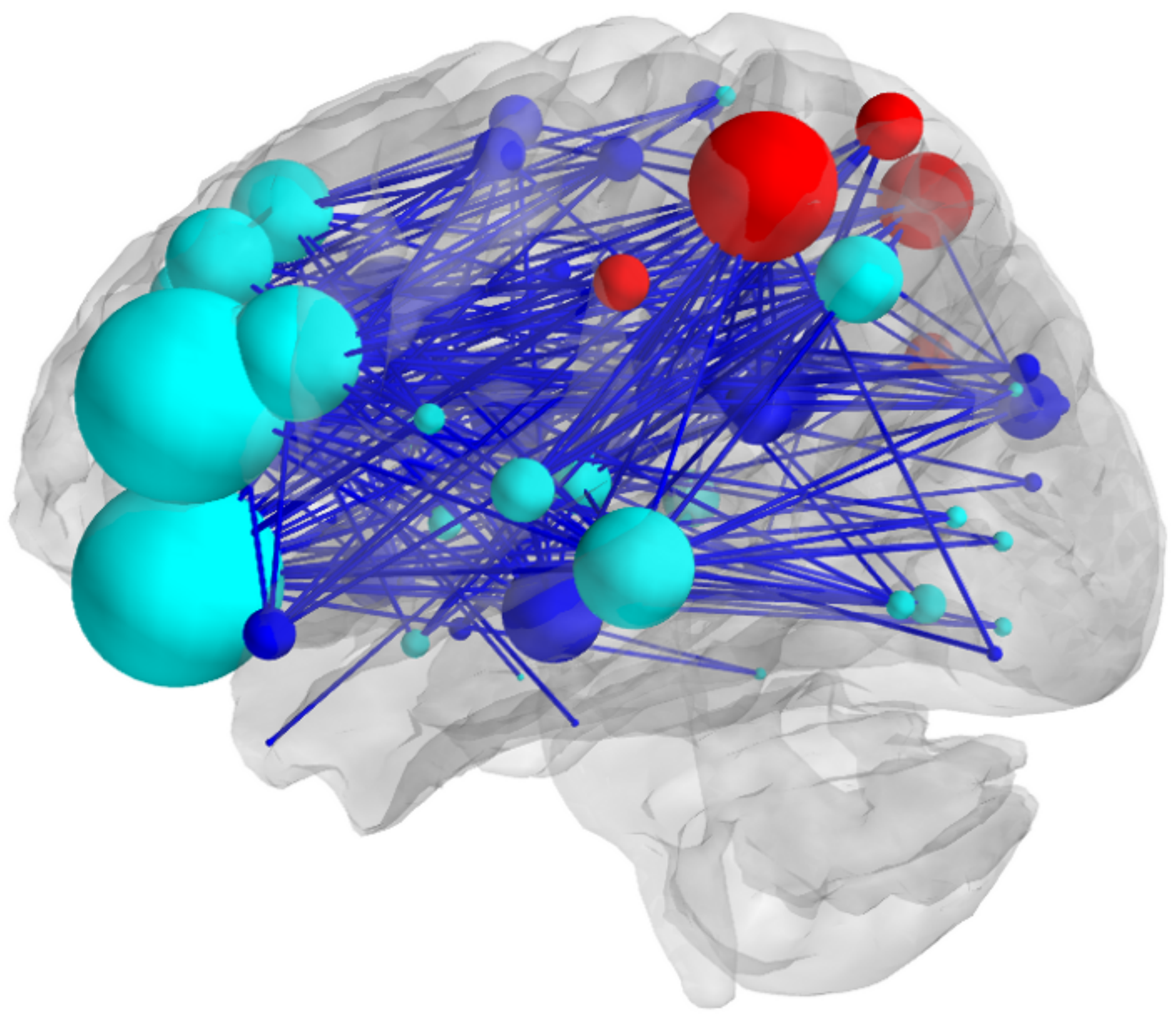}\\
   \tb{L} \hspace{8.0cm} \tb{R}
   \caption{\small 
    \textbf{Visualization of a differential SPN, summarizing the
      effect of a cognitive experimental factor.} Sagittal
    section of a negative differential
    SPN, which represents the significantly `lost' 
    edges, due to the $N$-back experimental factor. 
    The presence of an edge is determined by the thresholding of
    $p$-values at $.01$, uncorrected \citep[see][for a description of
    the data at hand.]{Ginestet2011a}.
    \label{fig:net_sagittal}}
\end{figure}

As for edge-specific differential SPNs, a vertex would be estimated to be either
significantly upweighted or downweighted, depending on the sign of the 
largest coefficient in the corresponding vector $\bbe^{v}$.
An illustration of such a differential SPN, based on the
$N$-back data set, analyzed by \citet{Ginestet2012a} is reported in figure
\ref{fig:net_sagittal}. Naturally, this assignment based on the sign of the
fixed effects is only possible, when the task under scrutiny is
based on an experimental gradient. An alternative strategy may be
required, when different levels of the task are expected to affect
the response in different directions.

A singular limitation, however, affects all mass-univariate
approaches. Such a repetitive use of classical inferential threshold,
may lead to a corresponding increase in Type I error. This issue
can be addressed by correcting for multiple comparisons. The
significance of edges and nodes in both mean
and differential SPNs can, for instance, be inferred using the false
discovery rate (FDR) with a base rate of $\alpha_{0}=.05$ 
\citep{Benjamini1995,Nichols2003}. Naturally, other corrections for
multiple comparisons could also be utilized \citep[see][for a
different approach]{Meskaldji2011}. The conventional thresholding
method used in network analysis is therefore superseded by the application of standard
multiple testing corrections. The main advantage of this approach lies
in its pooling of information over several subjects, in order to produce robust
edge- and node-specific statistics. 
%%%%%%%%%%%%%%%%%%%%%%%
%%%%%%%%%%%%%%%%%%%%%%%

%%%%%%%%%%%%%%%%%%%%%%%%%%%%%%%%%%
%%%%%%%%%%%%%%%%%%%%%%%%%%%%%%%%%%
%%%%%%%%%%%%%%%%%%%%%%%%%%%%%%%%%%
\section{Comparison of Functions on Networks}\label{sec:comparison}
We now turn to the issue of comparing various types of topological
measures over several families of networks \citep{Wijk2010}. Inference
on quantities such as characteristic path length, clustering coefficient and
modularity structure has attracted a sustained amount of interest in
the neuroscience community. Comparisons of this type of topological
measures, however, is generally regarded to be hard, since these
topological differences highly depend, in a non-linear fashion, on
group differences in edge density.

%%%%%%%%%%%%%%%%%%%%%%%%%%%%%%%%%%
\subsection{Global Efficiency}\label{sec:efficiency}
One of the classical exemplars of a topological summary of a network
is its characteristic path length. Such a quantity, however, is solely
defined for connected graphs. The global efficiency of a graph, by
contrast, can be computed for any network --connected or
disconnected-- and is inversely related to its characteristic path
length. Efficiency is formally defined by the following formula due to
\citet{Latora2001},
\begin{equation}
    E(G) = \frac{1}{N_{V}(N_{V}-1)}\sum_{i\in V}\sum_{j\neq i\in V} d^{-1}_{ij},
   \label{eq:general efficiency}  
\end{equation}
with $N_{V}=|V|$, as before. Here, $d_{ij}$ denotes the length of the shortest
path between vertices $i$ and $j$. Moreover, the second summation is
performed with respect to the set, $\lb j\neq i\in V\rb$, which 
is the set of all indices in $V$ that 
are different from $i$. This efficiency measure can be shown to be 
equivalent to the inverse of the harmonic mean of the length of the
shortest paths between each pair of nodes in the network $G$.

Specifically, the quantity in equation (\ref{eq:general efficiency}),
is usually referred to as the global efficiency of a particular graph,
and is denoted by $E^{\op{Glo}}(G) = E(G)$. Intuitively, this quantity can be
understood as the amount of potential information transfer that can be
performed in parallel. A local measure of efficiency can also be
computed, which is equivalent to the clustering coefficient. 
For a review of other efficiency measures that have been studied in the
context of neuroscience, the reader is referred to
\citet{Ginestet2011b}. The most commonly adopted approach to network
comparison is therefore to compute a topological metric, such as
global efficiency, for each individual subject, and thereafter to
evaluate whether this measure differs over the different experimental
groups under scrutiny. 

%%%%%%%%%%%%%%%%%%%%%%%%%%%%%%%%%%
\subsection{Density-integrated Measures}\label{sec:density-integrated}
An alternative approach to the problem of quantifying the topology of
weighted networks proceeds by integrating the metric of interest
with respect to different density levels. Different approaches have
been adopted in practice. While some authors have integrated over a subset
of the density range \citep[see][for example]{Achard2007}, others
have integrated over the entire range of densities \citep{He2009a}. The
family of topological measures, which is obtained after integrating over different
density levels, will be referred to as density-integrated
measures. Given a weighted graph $G=(\cV,\cE,\cW)$, the
density-integrated version of the efficiency in equation
(\ref{eq:general efficiency}) can, for instance, be defined as
follows,
\begin{equation}
     E_{K}(G) = \int\limits E(\gamma(G,k))p(k)dk,   
     \label{eq:density theoretical}
\end{equation}
where density is treated as a discrete random variable $K$, with realizations
in lower case, and $p(k)$ denotes the probability density 
function of $K$. Since $K$ is discrete, it can only take a countably finite
number of values. In general, it is common to assign equal weight to
every possible choice of density. 

The function $\gamma(G,k)$ in equation (\ref{eq:density theoretical})
is a density-thresholding function, which takes a weighted undirected
network and a level of wiring density as arguments, and returns an
\ti{unweighted} network. Since there is no prior knowledge about which
values of $K$ should be favored, one can specify a uniform distribution on
the set of all possible densities. Note, however, that other distributions could be
selected for this purpose \citep[see][for a discussion of
alternative specifications]{Ginestet2011b}. 

\subsection{Integrating over Densities}\label{sec:integrating density}
The question of separating topology from density could be reformulated
as the statistical problem of evaluating topological
differences, while `controlling' for differences in density. When adopting
this perspective, it is convenient to treat topology and density as
random variables. We have already done so for density, in the previous
section. Implicitly, by integrating over all possible thresholds, we
are indeed considering density as a random variable with a
well-defined probability distribution, which is, in the present case,
a uniform distribution. 

A natural desideratum, which may be required when comparing network
topological characteristics, while controlling for differences in
topology; would be to control for weighted networks whose association
matrices are proportional to each other. That is, if two different
matrices are linearly related to each other, it seems reasonable to
conclude that their topologies must be identical, after one has
controlled for such a linear difference in density. Thus, consider the
following simple example, adapted from \citet{Ginestet2011b}.

\begin{exa}[\textbf{\citet{Ginestet2011b}}]\label{exa:proportional2}
    We here have two networks, $G_{1}$ and $G_{2}$, with proportional
    association matrices $\bW_{1}$ and $\bW_{2}$, satisfying
    $\bW_{1}=\alpha\bW_{2}$. That is, these two matrices are
    proportional to each other. 
    An application of the density-integrated metrics described in equation 
    (\ref{eq:density theoretical}) to these networks would 
    give the following equalities, 
    \begin{equation}
           E_{K}(\bW_{1}) = E_{K}(\alpha\bW_{2}) = E_{K}(\bW_{2}). 
    \end{equation}
    That is, when integrating with respect to density, we are in fact evaluating the
    efficiencies of $G_{1}$ and $G_{2}$ at a number of cut-off 
    points. At each of these points, the efficiency of the two networks will
    be identical, because $\bW_{1}$ is proportional to $\bW_{2}$ and
    therefore the same sets of edges will be selected. 
    Therefore, $G_{1}$ and $G_{2}$ have identical density-integrated
    efficiencies. 
\end{exa}

While illustrative, this example is not entirely satisfying. In fact,
this result can be shown to hold in a more general sense. The invariance of 
density-integrated efficiency turns out to be true for any monotonic
(increasing or decreasing) function $h$, as formally stated in the
following result.
\begin{pro}[\textbf{\citet{Ginestet2011b}}]\label{pro:monotonic}
   Let a weighted undirected graph $G=(\cV,\cE,\cW)$. 
   For any monotonic function $h(\cdot)$ acting elementwise on a
   real-valued matrix $\bW$, and any topological metric $E$,
   the density-integrated version of that metric, denoted $E_{K}$, satisfies
   \begin{equation}
          E_{K}(\cW) = E_{K}(h(\cW)),
   \end{equation}
   where we have used the weight set, $\cW$, as a proxy for graph $G$.
\end{pro}
A proof of this proposition can be found in \citet{Ginestet2011b}.
The demonstration essentially relies on the fact that any monotonic
transformation of the entries of a real-valued matrix will preserve
the ranks. Therefore, proposition \ref{pro:monotonic} makes rigorous
a potential way of ``controlling'' for differences in density. That
is, this formal proposition states that
we are indeed controlling for any monotonic transformation of the original
entries in the matrix. In effect, proposition \ref{pro:monotonic}
should be regarded as a potential definition of what it means for two
networks to solely differ in terms of topology, while controlling for
monotonic differences in density. 

%%%%%%%%%%%%%%%%%%%%%%%%%%%%%%%%%%
%%%%%%%%%%%%%%%%%%%%%%%%%%%%%%%%%%
%%%%%%%%%%%%%%%%%%%%%%%%%%%%%%%%%%
\subsection{Density and Modularity}\label{sec:modularity}
Another network property, which has been studied extensively in the
literature is modularity structure. As for efficiency and other
topological measures, however, modularity is also highly dependent on
edge density. Therefore, any attempt at comparing the modularity of
different groups of networks will be confounded by group differences
in the networks' number of edges. We illustrate this problem with the
results reported in a recent paper by \citet{Bassett2011}, who have
analyzed the static and
dynamic organization of functional brain networks in humans. We here focus on the
first claim made in this paper, which states that the static modular
structure of such networks is nested with respect to time. 
In particular, \citet{Bassett2011} argue that this graded structure
underlines a ``multiscale modular structure''. 

As for global efficiency in the previous section, it can be shown that
modularity structure is substantially mediated by edge density. In
the case of weighted networks, this is equivalent to a difference in the
size of the correlation coefficients. In \citet{Bassett2011}, for
instance, the authors report that the size of the mean correlation
diminishes with the size of the time window. Such a decrease in overall correlation 
will generally have two effects: (i) networks'
topologies will become increasingly more ``random'' and (ii) the number of
significant edges will decrease. Here, we use synthetic
data sets to show that these two phenomena are likely to be associated
with a higher number of modules, thereby potentially explaining the apparent
multiscale modular structure described by \citet{Bassett2011}. Our
simulations are based on the unweighted unsigned version of the
modularity algorithm of \citet{Clauset2004}, but may be extrapolated
to weighted signed adjacency matrices.
%%%%%%%%%%%%%%%%%%%%%%%%%%%%%%%%%%
\begin{figure}[htbp]  
  {\Large\tb{A}}\hspace{6.3cm}{\Large\tb{B}}\\ 
  \includegraphics{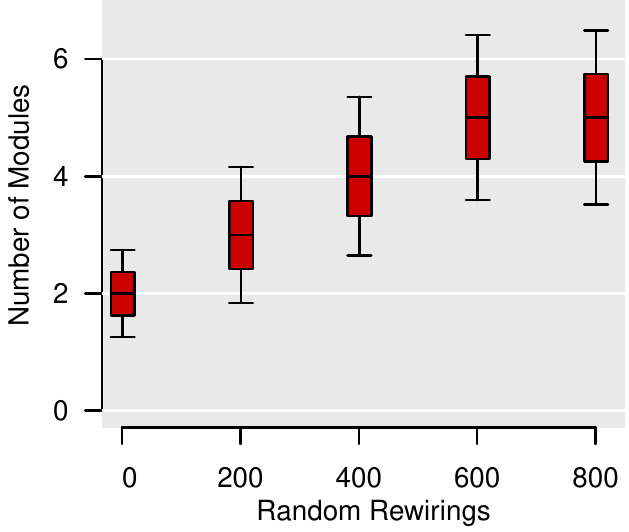}
  \hspace{.3cm}              
  \includegraphics{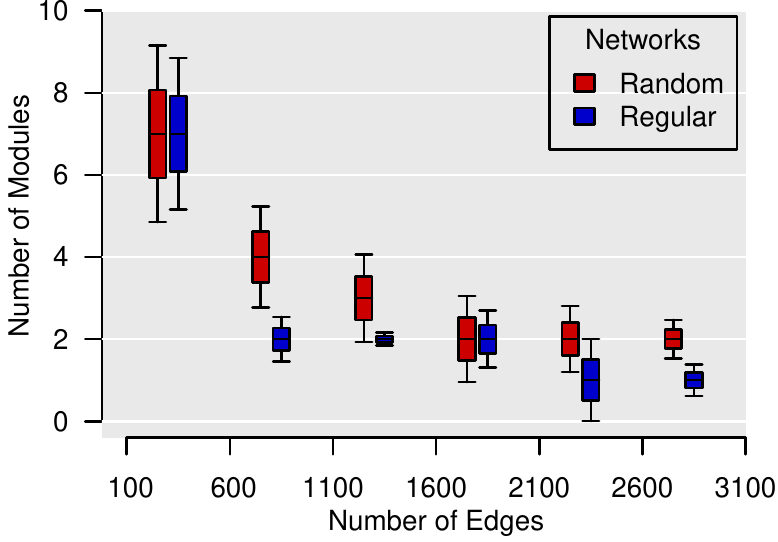} 
  \\
  {\Large\tb{C}}\vspace{-.5cm}
  \begin{center}
    \includegraphics[width=15cm]{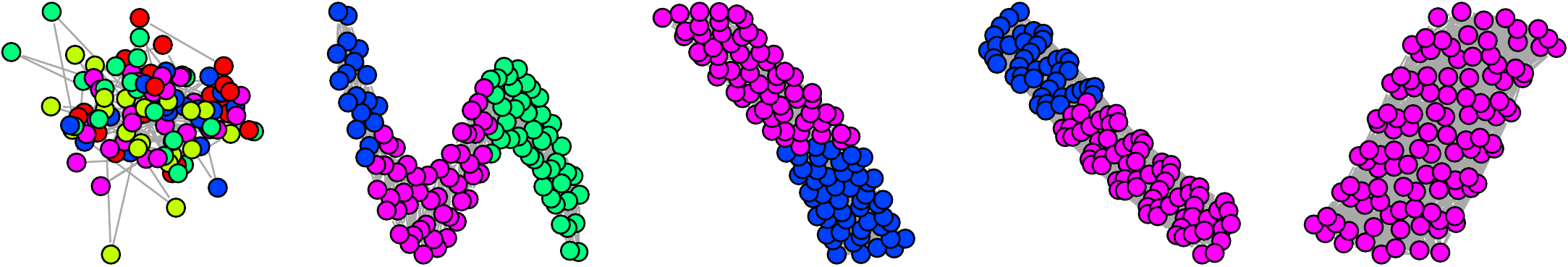}
  {\footnotesize
  \tb{$N_{E}=100$}\hspace{1.5cm}
  \tb{$N_{E}=600$}\hspace{1.5cm}
  \tb{$N_{E}=1100$}\hspace{1.5cm}
  \tb{$N_{E}=1600$}\hspace{1.5cm}
  \tb{$N_{E}=2100$}}
  \end{center}
  \vspace{.5cm}
  {\Large\tb{D}}
  \vspace{-.5cm}
  \begin{center}
    \includegraphics[width=15cm]{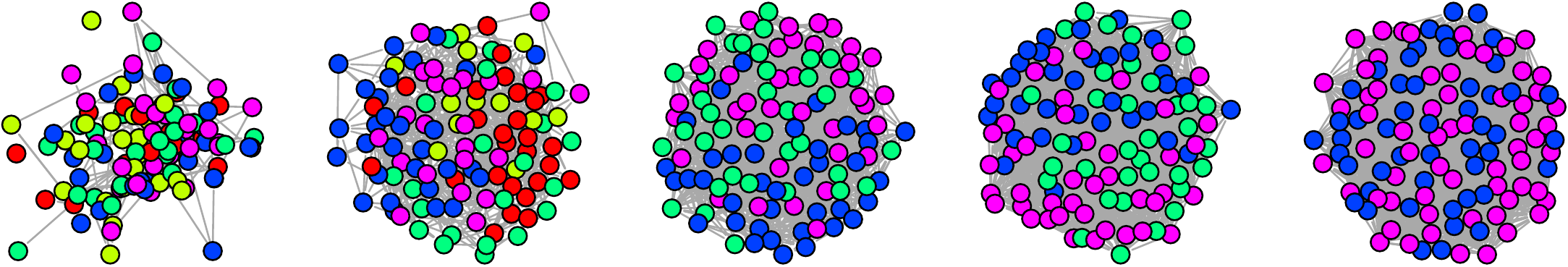}
  {\footnotesize
  \tb{$N_{E}=100$}\hspace{1.5cm}
  \tb{$N_{E}=600$}\hspace{1.5cm}
  \tb{$N_{E}=1100$}\hspace{1.5cm}
  \tb{$N_{E}=1600$}\hspace{1.5cm}
  \tb{$N_{E}=2100$}}
  \end{center}
  \caption{\tb{Topological randomness and number of edges predict number of
      modules.} (A) Relationship between the number of random rewirings
    of a regular lattice and the number of modules in such a
    network. Here, the number of edges is kept constant throughout all
    rewirings. (B) Relationship between the number of edges in a
    network and its number of modules for both regular (i.e. lattice) and random
    graphs. This shows that the number of modules tends to decrease as
    more edges are added to both types of networks. (C-D)
    Modular structures of regular (C) and random (D) networks for
    different number of edges, $N_{E}$. These networks are represented
    using the algorithm of 
    \citet{Kamada1989} with different colors representing different
    modules. In all simulations, the number of vertices is
    $N_{V}=112$, as in \citet{Bassett2011}.
    \label{fig:nets}}
\end{figure}
%%%%%%%%%%%%%%%%%%%%%%%%%%%%%%%%%%

In panel (A) of figure \ref{fig:nets}, we have generated 1,000 unweighted
lattices based on 112 vertices as in \citet{Bassett2011}. By randomly
rewiring the edges of these lattices, we show that 
the number of modules in these networks tends to increase with the
level of topological randomness in these graphs.
For panels (B) to (D), we have generated two sets
of unweighted networks, characterized by a random and a regular
topology, respectively, with different number of edges. 
These simulations were repeated 1,000 times for each type of graph for
each number of edges. For  both types of networks, the number of
modules in these graphs tended to decrease as new edges were added. 
Collectively, although these data simulations do not entirely rule out
the possibility of a temporally nested modular structure in the human brain,
they nonetheless cast doubts on the possibility of detecting such a temporal
organization by reducing the size of the sampling window. Such subtle
artifactual relationships between modularity and edge density can 
arise in a range of different settings in the analysis of neuroimaging
data. 

%%%%%%%%%%%%%%%%%%%%%%%%%%%%%%%%%%
\subsection{Weighted Topological Metrics}\label{sec:weighted}
Since the previous two sections have highlighted the potential
problems associated with thresholding correlation matrices, one may
surmise that such problems could be adequately dealt with, by
directly considering the weighted versions of the topological metrics
of interest. In particular, an apparently natural way of combining
differences in density with differences in topology is to consider the
weighted versions of traditional topological metrics. For the
aforementioned global efficiency, for instance, one can define a weighted global
efficiency, denoted $E_{W}$, as follows, 
\begin{equation}
     E_{W}(G) = \frac{1}{N_{V}(N_{V}-1)}\sum_{i=1}^{N_{V}}\sum_{j\neq
    i}^{N_{V}} \frac{1}{d^{W}_{ij}} = \frac{1}{N_{I}}\sum_{\cI(G)} \frac{1}{d^{W}_{ij}},
    \label{eq:weighted definition}
\end{equation}
where $d^{W}_{ij}$ represents the weighted shortest path between the
$i\tth$ and $j\tth$ nodes. 
Unfortunately, another theoretical result points to a serious
limitation of $E_{W}$, which may
potentially dissuade researchers from using this particular type of metrics.
With the next proposition, we demonstrate that under mild
conditions, the weighted efficiency is simply equivalent to the
weighted density, sometimes referred to as weighted cost, of the graph of interest, 
\begin{equation}
   K_{W}(G) =  \frac{1}{N_{V}(N_{V}-1)}\sum_{i=1}^{N_{V}}\sum_{j\neq i}^{N_{V}} w_{ij}.
           \label{eq:weighted cost}
\end{equation}
%%%%%%%%%%%%%%%%%%%%%
\begin{pro}[\textbf{\citet{Ginestet2011b}}]\label{pro:weighted-density}
   For any weighted graph $G=(\cV,\cE,\cW)$, whose weighted edge set is
   denoted by $\cW(G)=\lb w_{ij}:i<j \rb$, if 
   \begin{equation}
        \min_{w_{ij}\in\cW(G)} w_{ij} \geq \frac{1}{2}\max_{w_{ij}\in\cW(G)}w_{ij},
   \end{equation}
   then 
   \begin{equation}
          E_{W}(G) = K_{W}(G).
   \end{equation}
\end{pro}
%%%%%%%%%%%%%%%%%%%%%
A proof of this result can be found in \citet{Ginestet2011b}.
Not surprisingly, proposition \ref{pro:weighted-density} places
emphasis on the spread of the distribution of the weighted edge set $\cE(G)$. 
The condition in proposition \ref{pro:weighted-density} may at first
appear quite constraining. However, this condition encompasses a wide
range of experimental situations, including the data set described in  
\citet{Ginestet2011a}. Thus, the added
benefit of utilizing the weighted version of the global efficiency
measure may, in most settings, be highly questionable, since there
exists a one-to-one relationship between this
topological measure and a simple average of the edge
weights. Cutoff-integrated efficiency and other cutoff-integrated
measures, as described in \citet{Ginestet2011b}, may therefore be
preferred, in practice, when one wishes to
summarize the influence of both density and topological differences.
 
%%%%%%%%%%%%%%%%%%%%%%%%%%%%%%%%%%
%%%%%%%%%%%%%%%%%%%%%%%%%%%%%%%%%%
%%%%%%%%%%%%%%%%%%%%%%%%%%%%%%%%%%
\section{Conclusion}\label{sec:discussion}
In this paper, we have briefly reviewed some of the methodological
research that has been conducted on network data analysis, as applied
to functional neuroimaging. Two main threads ran through this
discussion. Firstly, we considered the different approaches that
one may adopt, when summarizing several subject-specific
networks. Secondly, the thorny issue of graph thresholding was
also tackled, with special emphasis on the comparison of network
modularity and the use of weighted topological metrics. 

From the above discussion, it should be clear that there does not
exist a single way of computing a mean network. This is, in some
sense, an ill-defined problem. A commonly adopted perspective on this issue is
to perform a mass-univariate test, where the significance levels of
every edge are evaluated, and then thresholded. We have seen that this
approach can be carried out both within a single family of networks,
and over an entire experimental design, using a mixed effects model. 
By analogy with the classical SPM approach used in neuroimaging, one
may refer to such uses of a mass-univariate approach on networks, as
SPNs. 

Secondly, we have discussed one of the long-standing issues in the
application of network data analysis to neuroscience data: the
question of whether or not one should threshold matrices of correlation
coefficients, for the purpose of producing adjacency matrices. In this
paper, we have reviewed a range of different approaches to this
problem. On the basis of the several 
examples and counterexamples that we have studied, we are able to make
a few methodological recommendations to researchers in the
neuroscience community, intending to compare the topological
properties of two or more populations of weighted networks. Note that
these recommendations are solely tentative, as no general consensus
has yet been reached on this particular issue. 

%Recommendations
As a first step, we argue that it is good practice to standardize the
association weights. This may facilitate comparison
across distinct network analyses, and ease the interpretation of the
results. Secondly, the weighted density, or connectivity strength,
of the networks of interest should then be reported. This is central to the rest
of the analysis, and therefore, this quantity should be computed and
reported systematically. Indeed, if the groups of networks under
scrutiny substantially differ in terms of average density, then these
differences are highly likely to affect any comparison of the
topological properties of these groups of networks. 
Finally, population differences in density-integrated topological metrics may then
be evaluated and reported. This will indicate whether the topologies of the 
populations under scrutiny vary significantly after having controlled
for monotonic differences in connectivity strength. 

% 2. 
The theoretical results described in this paper have only been presented for the
global efficiency metric. Thus, these propositions and the 
examples studied need not necessarily apply to other topological
measures. However, we also note that proposition \ref{pro:monotonic}
has been proved with a high degree of generality. This proposition and
its proof is indeed independent of the particular formula of the
metric of interest, and therefore could easily be extended to any
other function of the weighted graph matrix. In particular, because
most weighted metrics are constructed on the basis of the matrix of
weighted shortest paths, one surmises that this theoretical
result may, in fact, hold in a more general setting. 

% Modularity is not immune.
Importantly, we have also shown that network modularity is not
immune to this dependency on edge density. If several populations of
networks differ in their number of edges, then it is likely that the
resulting group-specific modularity structures will not be
comparable. That is, such comparisons will mainly reflect differences
in edge density, and as such may not carry much explanatory
power. This is an area of application of statical network analysis,
where one should exert caution, as the powerful algorithms used for
detecting network modules may hide the potential confounding effects
of differences in edge density. 

% Weighted metrics do not solve the problem.
Finally, the use of weighted topological metrics was also
considered. Unfortunately, we have seen that simply replacing
classical network measures by their weighted analogs is not
sufficient to resolve the dependency of these measures on edge
density. Thus, cutoff-integrated topological measures, such as the
cutoff-integrated efficiency described in \citet{Ginestet2011b}, may be
preferred in practice, when one wishes to separate differences in edge
density from differences in topology. 

% references --------------------------------------------------
% \footnotesize
\singlespacing
\addcontentsline{toc}{section}{References}
\bibliography{/home/cgineste/ref/bibtex/Statistics,%
              /home/cgineste/ref/bibtex/Neuroscience}

\begin{thebibliography}{55}
\providecommand{\natexlab}[1]{#1}

\bibitem[{Achard and Bullmore(2007)}]{Achard2007}
Achard, S. and Bullmore, E. (2007).
\newblock Efficiency and cost of economical brain functional networks.
\newblock \textit{PLOS Computational Biology}, \textbf{3}, 174--182.

\bibitem[{Achard et~al.(2006)Achard, Salvador, Whitcher, Suckling, and
  Bullmore}]{Achard2006}
Achard, S., Salvador, R., Whitcher, B., Suckling, J., and Bullmore, E. (2006).
\newblock A resilient, low-frequency, small-world human brain functional
  network with highly connected association cortical hubs.
\newblock \textit{J. Neurosci.}, \textbf{26}(1), 63--72.

\bibitem[{Astolfi et~al.(2009)Astolfi, Cincotti, Mattia, De~Vico~Fallani,
  Salinari, Marciani, Witte, and Babiloni}]{Astolfi2009}
Astolfi, L., Cincotti, F., Mattia, D., De~Vico~Fallani, F., Salinari, S.,
  Marciani, M., Witte, H., and Babiloni, F. (2009).
\newblock Study of the time-varying cortical connectivity changes during the
  attempt of foot movements by spinal cord injured and healthy subjects.
\newblock \textit{Conf Proc IEEE Eng Med Biol Soc}, 2208--11.

\bibitem[{Balding et~al.(2009)Balding, Ferrari, Fraiman, and
  Sued}]{Balding2009}
Balding, D., Ferrari, P., Fraiman, R., and Sued, M. (2009).
\newblock Limit theorems for sequences of random trees.
\newblock \textit{TEST}, \textbf{18}(2), 302--315.

\bibitem[{Barabasi and Albert(1999)}]{Barabasi1999}
Barabasi, A.L. and Albert, R. (1999).
\newblock Emergence of scaling in random networks.
\newblock \textit{Science}, \textbf{286}, 509--512.

\bibitem[{Bartlett(1937)}]{Bartlett1937}
Bartlett, M.S. (1937).
\newblock Properties of sufficiency and statistical tests.
\newblock \textit{Proceedings of the Royal Society of London. Series A,
  Mathematical and Physical Sciences}, \textbf{160}(901), 268--282.

\bibitem[{Bassett et~al.(2008)Bassett, Bullmore, Verchinski, Mattay,
  Weinberger, and Meyer-Lindenberg}]{Bassett2008}
Bassett, D.S., Bullmore, E., Verchinski, B.A., Mattay, V.S., Weinberger, D.R.,
  and Meyer-Lindenberg, A. (2008).
\newblock Hierarchical organization of human cortical networks in health and
  schizophrenia.
\newblock \textit{J. Neurosci.}, \textbf{28}(37), 9239--9248.

\bibitem[{Bassett et~al.(2006)Bassett, Meyer-Lindenberg, Achard, Duke, and
  Bullmore}]{Bassett2006}
Bassett, D.S., Meyer-Lindenberg, A., Achard, S., Duke, T., and Bullmore, E.
  (2006).
\newblock Adaptive reconfiguration of fractal small-world human brain
  functional networks.
\newblock \textit{Proceedings of the National Academy of Sciences of the United
  States of America}, \textbf{103}(51), 19518--19523.

\bibitem[{Bassett et~al.(2011)Bassett, Wymbs, Porter, Mucha, Carlson, and
  Grafton}]{Bassett2011}
Bassett, D.S., Wymbs, N.F., Porter, M.A., Mucha, P.J., Carlson, J.M., and
  Grafton, S.T. (2011).
\newblock Dynamic reconfiguration of human brain networks during learning.
\newblock \textit{Proceedings of the National Academy of Sciences},
  \textbf{108}(18), 7641--7646.

\bibitem[{Benjamini and Hochberg(1995)}]{Benjamini1995}
Benjamini, Y. and Hochberg, Y. (1995).
\newblock Controlling the false discovery rate: A practical and powerful
  approach to multiple testing.
\newblock \textit{Journal of the Royal Statistical Society. Series B
  (Methodological)}, \textbf{57}(1), 289--300.

\bibitem[{Bigot and Charlier(2011)}]{Bigot2011}
Bigot, J. and Charlier, B. (2011).
\newblock On the consistency of {F}rechet means in deformable models for curve
  and image analysis.
\newblock \textit{Electronic Journal of Statistics}, \textbf{5}, 1054--1089.

\bibitem[{Billingsley(1995)}]{Billingsley1995}
Billingsley, P. (1995).
\newblock \textit{Probability and Measure.}
\newblock Wiley Series in Probability and Mathematical Statistics. Wiley, New
  York.

\bibitem[{Bollob\'{a}s(1998)}]{Bollobas1998}
Bollob\'{a}s, B. (1998).
\newblock \textit{Modern Graph Theory.}
\newblock Springer, London.

\bibitem[{Bowman(2007)}]{Bowman2007}
Bowman, F.D. (2007).
\newblock Spatiotemporal models for region of interest analyses of functional
  neuroimaging data.
\newblock \textit{Journal of the American Statistical Association},
  \textbf{102}(478), 442--453.

\bibitem[{Bullmore and Sporns(2009)}]{Bullmore2009}
Bullmore, E. and Sporns, O. (2009).
\newblock Complex brain networks: Graph theoretical analysis of structural and
  functional systems.
\newblock \textit{Nature Reviews Neuroscience}, \textbf{10(1)}, 1--13.

\bibitem[{Cecchi et~al.(2007)Cecchi, Rao, Centeno, Baliki, Apkarian, and
  Chialvo}]{Cecchi2007}
Cecchi, G., Rao, A., Centeno, M., Baliki, M., Apkarian, A., and Chialvo, D.
  (2007).
\newblock Identifying directed links in large scale functional networks:
  application to brain f{MRI}.
\newblock \textit{BMC Cell Biology}, \textbf{8}, 1--10.

\bibitem[{Clauset et~al.(2004)Clauset, Newman, and Moore}]{Clauset2004}
Clauset, A., Newman, M.E.J., and Moore, C. (2004).
\newblock Finding community structure in very large networks.
\newblock \textit{Phys. Rev. E}, \textbf{70}(6), 066111--.

\bibitem[{De~Vico~Fallani et~al.(2008)De~Vico~Fallani, Astolfi, Cincotti,
  Mattia, Marciani, Tocci, Salinari, Witte, Hesse, Gao, Colosimo, and
  Babiloni}]{Fallani2008}
De~Vico~Fallani, F., Astolfi, L., Cincotti, F., Mattia, D., Marciani, M.,
  Tocci, A., Salinari, S., Witte, H., Hesse, W., Gao, S., Colosimo, A., and
  Babiloni, F. (2008).
\newblock Cortical network dynamics during foot movements.
\newblock \textit{Neuroinformatics}, \textbf{6}(1), 23--34.

\bibitem[{Demidenko(2004)}]{Demidenko2004}
Demidenko, E. (2004).
\newblock \textit{Mixed Models: Theory and Applications}.
\newblock Wiley, London.

\bibitem[{Dryden and Mardia(1998)}]{Dryden1998}
Dryden, I. and Mardia, K. (1998).
\newblock \textit{Statistical Analysis of Shape}.
\newblock Wiley, London.

\bibitem[{Fournel et~al.(2013)Fournel, Reynaud, Brammer, Simmons, and
  Ginestet}]{Fournel2013}
Fournel, A.P., Reynaud, E., Brammer, M., Simmons, A., and Ginestet, C.E.
  (2013).
\newblock Group analysis of self-organizing maps based on functional mri using
  restricted frechet means.
\newblock \textit{Neuroimage}, \textbf{76}, 373--385.

\bibitem[{Fr\'{e}chet(1948)}]{Frechet1948}
Fr\'{e}chet, M. (1948).
\newblock Les \'{e}l\'{e}ments al\'{e}atoires de nature quelconque dans un
  espace distanci\'{e}.
\newblock \textit{Annales de L'{I}nstitut {H}enri {P}oincar\'{e}},
  \textbf{10(4)}, 215--310.

\bibitem[{Friston(1994)}]{Friston1994}
Friston, K.J. (1994).
\newblock Functional and effective connectivity in neuroimaging: A synthesis.
\newblock \textit{Human Brain Mapping}, \textbf{2}(1-2), 56--78.

\bibitem[{Ginestet et~al.(2012{\natexlab{a}})Ginestet, Best, and
  Richardson}]{Ginestet2012a}
Ginestet, C.E., Best, N.G., and Richardson, S. (2012{\natexlab{a}}).
\newblock Classification loss function for parameter ensembles in bayesian
  hierarchical models.
\newblock \textit{Statistics and Probability Letters}, \textbf{82}(0),
  859--863.

\bibitem[{Ginestet et~al.(2011)Ginestet, Nichols, Bullmore, and
  Simmons}]{Ginestet2011b}
Ginestet, C.E., Nichols, T.E., Bullmore, E.T., and Simmons, A. (2011).
\newblock Brain network analysis: Separating cost from topology using
  cost-integration.
\newblock \textit{PLoS ONE}, \textbf{6(7)}, e21570.

\bibitem[{Ginestet and Simmons(2011)}]{Ginestet2011a}
Ginestet, C.E. and Simmons, A. (2011).
\newblock Statistical parametric network analysis of functional connectivity
  dynamics during a working memory task.
\newblock \textit{NeuroImage}, \textbf{5(2)}, 688--704.

\bibitem[{Ginestet et~al.(2012{\natexlab{b}})Ginestet, Simmons, and
  Kolaczyk}]{Ginestet2012b}
Ginestet, C.E., Simmons, A., and Kolaczyk, E.D. (2012{\natexlab{b}}).
\newblock Weighted frechet means as convex combinations in metric spaces:
  Properties and generalized median inequalities.
\newblock \textit{Statistics and Probability Letters}, \textbf{82}(10),
  1859--1863.

\bibitem[{Gong et~al.(2009)Gong, He, Concha, Lebel, Gross, Evans, and
  Beaulieu}]{Gong2009}
Gong, G., He, Y., Concha, L., Lebel, C., Gross, D.W., Evans, A.C., and
  Beaulieu, C. (2009).
\newblock Mapping anatomical connectivity patterns of human cerebral cortex
  using in vivo diffusion tensor imaging tractography.
\newblock \textit{Cereb. Cortex}, \textbf{19}(3), 524--536.

\bibitem[{Hagmann et~al.(2008)Hagmann, Cammoun, Gigandet, Meuli, Honey, Wedeen,
  and Sporns}]{Hagmann2008}
Hagmann, P., Cammoun, L., Gigandet, X., Meuli, R., Honey, C.J., Wedeen, V.J.,
  and Sporns, O. (2008).
\newblock Mapping the structural core of human cerebral cortex.
\newblock \textit{PLoS Biol}, \textbf{6}(7), 159--169.

\bibitem[{He et~al.(2007)He, Chen, and Evans}]{He2007}
He, Y., Chen, Z.J., and Evans, A.C. (2007).
\newblock Small-world anatomical networks in the human brain revealed by
  cortical thickness from {MRI}.
\newblock \textit{Cereb. Cortex}, \textbf{17}(10), 2407--2419.

\bibitem[{He et~al.(2009{\natexlab{a}})He, Dagher, Chen, Charil, Zijdenbos,
  Worsley, and Evans}]{He2009a}
He, Y., Dagher, A., Chen, Z., Charil, A., Zijdenbos, A., Worsley, K., and
  Evans, A. (2009{\natexlab{a}}).
\newblock Impaired small-world efficiency in structural cortical networks in
  multiple sclerosis associated with white matter lesion load.
\newblock \textit{Brain}, \textbf{132}(12), 3366--3379.

\bibitem[{He et~al.(2009{\natexlab{b}})He, Wang, Wang, Chen, Yan, Yang, Tang,
  Zhu, Gong, Zang, and Evans}]{He2009}
He, Y., Wang, J., Wang, L., Chen, Z.J., Yan, C., Yang, H., Tang, H., Zhu, C.,
  Gong, Q., Zang, Y., and Evans, A.C. (2009{\natexlab{b}}).
\newblock Uncovering intrinsic modular organization of spontaneous brain
  activity in humans.
\newblock \textit{PLoS ONE}, \textbf{4}(4), 1--18.

\bibitem[{Kamada and Kawai(1989)}]{Kamada1989}
Kamada, T. and Kawai, S. (1989).
\newblock An algorithm for drawing general undirected graphs.
\newblock \textit{Information Processing Letters}, \textbf{31(1)}, 7--15.

\bibitem[{Karcher(1977)}]{Karcher1977}
Karcher, H. (1977).
\newblock Riemannian center of mass and mollifier smoothing.
\newblock \textit{Communication in Pure and Applied Mathematics}, \textbf{30},
  509--541.

\bibitem[{Kolaczyk(2009)}]{Kolaczyk2009}
Kolaczyk, E. (2009).
\newblock \textit{Statistical Analysis of Network Data: Methods and Models}.
\newblock Springer-Verlag, London.

\bibitem[{Laird and Ware(1982)}]{Laird1982}
Laird, N. and Ware, J. (1982).
\newblock Random-effects models for longitudinal data.
\newblock \textit{Biometrics}, \textbf{38}, 963--974.

\bibitem[{Latora and Marchiori(2001)}]{Latora2001}
Latora, V. and Marchiori, M. (2001).
\newblock Efficient behavior of small-world networks.
\newblock \textit{Phys. Rev. Lett.}, \textbf{87}(19), 198701--198705.

\bibitem[{Lin et~al.(2006)Lin, Louis, Paddock, and Ridgeway}]{Lin2006}
Lin, R., Louis, T., Paddock, S., and Ridgeway, G. (2006).
\newblock Loss function based ranking in two-stage hierarchical models.
\newblock \textit{Bayesian analysis}, \textbf{1(4)}, 915--946.

\bibitem[{Meskaldji et~al.(2011)Meskaldji, Ottet, Cammoun, Hagmann, Meuli,
  Eliez, Thiran, and Morgenthaler}]{Meskaldji2011}
Meskaldji, D.E., Ottet, M.C., Cammoun, L., Hagmann, P., Meuli, R., Eliez, S.,
  Thiran, J.P., and Morgenthaler, S. (2011).
\newblock Adaptive strategy for the statistical analysis of connectomes.
\newblock \textit{PloS one}, \textbf{6}(8), e23009--.

\bibitem[{Nichols and Hayasaka(2003)}]{Nichols2003}
Nichols, T. and Hayasaka, S. (2003).
\newblock Controlling the familywise error rate in functional neuroimaging: a
  comparative review.
\newblock \textit{Statistical Methods in Medical Research}, 419--446.

\bibitem[{Pachou et~al.(2008)Pachou, Vourkas, Simos, Smit, Stam, Tsirka, and
  Micheloyannis}]{Pachou2008}
Pachou, E., Vourkas, M., Simos, P., Smit, D., Stam, C., Tsirka, V., and
  Micheloyannis, S. (2008).
\newblock Working memory in schizophrenia: An {EEG} study using power spectrum
  and coherence analysis to estimate cortical activation and network behavior.
\newblock \textit{Brain Topography}, \textbf{21}(2), 128--137.

\bibitem[{Parthasarathy(1967)}]{Parthasarathy1967}
Parthasarathy, K. (1967).
\newblock \textit{Probability Measures on Metric Spaces}.
\newblock American Mathematical Society, London.

\bibitem[{Pinheiro and Bates(2000)}]{Pinheiro2000}
Pinheiro, J. and Bates, D. (2000).
\newblock \textit{Mixed-effects mode in S and S-Plus}.
\newblock Springer, London.

\bibitem[{Richiardi et~al.(2011)Richiardi, Eryilmaz, Schwartz, Vuilleumier, and
  Van De~Ville}]{Richiardi2011}
Richiardi, J., Eryilmaz, H., Schwartz, S., Vuilleumier, P., and Van De~Ville,
  D. (2011).
\newblock Decoding brain states from f{MRI} connectivity graphs.
\newblock \textit{NeuroImage}, \textbf{56}, 616--626.

\bibitem[{Rubinov and Sporns(2010)}]{Rubinov2010}
Rubinov, M. and Sporns, O. (2010).
\newblock Complex network measures of brain connectivity: Uses and
  interpretations.
\newblock \textit{Neuroimage}, \textbf{52}, 1059--1069.

\bibitem[{Salvador et~al.(2008)Salvador, Martinez, Pomarol-Clotet, Gomar, Vila,
  Sarro, Capdevila, and Bullmore}]{Salvador2008}
Salvador, R., Martinez, A., Pomarol-Clotet, E., Gomar, J., Vila, F., Sarro, S.,
  Capdevila, A., and Bullmore, E. (2008).
\newblock A simple view of the brain through a frequency-specific functional
  connectivity measure.
\newblock \textit{NeuroImage}, \textbf{39}(1), 279--289.

\bibitem[{Searc\'{o}id(2007)}]{Searcoid2007}
Searc\'{o}id, M. (2007).
\newblock \textit{Metric spaces.}
\newblock Springer, London.

\bibitem[{Snijders et~al.(2010{\natexlab{a}})Snijders, Koskinen, and
  Schweinberger}]{Snijders2010a}
Snijders, T.A., Koskinen, J., and Schweinberger, M. (2010{\natexlab{a}}).
\newblock Maximum likelihood estimation for social network dynamics.
\newblock \textit{The Annals of Applied Statistics}, \textbf{4}(2), 567--588.

\bibitem[{Snijders et~al.(2010{\natexlab{b}})Snijders, Van~de Bunt, and
  Steglich}]{Snijders2010}
Snijders, T.A., Van~de Bunt, G.G., and Steglich, C.E. (2010{\natexlab{b}}).
\newblock Introduction to stochastic actor-based models for network dynamics.
\newblock \textit{Social networks}, \textbf{32}(1), 44--60.

\bibitem[{Sverdrup-Thygeson(1981)}]{Sverdrup1981}
Sverdrup-Thygeson, H. (1981).
\newblock Strong law of large numbers for measures of central tendency and
  dispersion of random variables in compact metric spaces.
\newblock \textit{The Annals of Statistics}, \textbf{9}(1), 141--145.

\bibitem[{Thorstensen et~al.(2009)Thorstensen, Segonne, and
  Keriven}]{Thorstensen2009}
Thorstensen, N., Segonne, F., and Keriven, R. (2009).
\newblock \textit{Scale Space and Variational Methods in Computer Vision}, vol.
  5567, chap. Pre-image as Karcher Mean Using Diffusion Maps: Application to
  Shape and Image Denoising, 721--732.
\newblock Springer.

\bibitem[{van~den Heuvel et~al.(2009)van~den Heuvel, Stam, Kahn, and
  Hulshoff~Pol}]{Heuvel2009}
van~den Heuvel, M.P., Stam, C.J., Kahn, R.S., and Hulshoff~Pol, H.E. (2009).
\newblock Efficiency of functional brain networks and intellectual performance.
\newblock \textit{J. Neurosci.}, \textbf{29}(23), 7619--7624.

\bibitem[{van Wijk et~al.(2010)van Wijk, Stam, and Daffertshofer}]{Wijk2010}
van Wijk, B.C.M., Stam, C.J., and Daffertshofer, A. (2010).
\newblock Comparing brain networks of different size and connectivity density
  using graph theory.
\newblock \textit{PLoS ONE}, \textbf{5}(10), 13701--13716.

\bibitem[{Watts and Strogatz(1998)}]{Watts1998}
Watts, D.J. and Strogatz, S.H. (1998).
\newblock Collective dynamics of `small-world' networks.
\newblock \textit{Nature}, \textbf{393}(6684), 440--442.

\bibitem[{Zalesky et~al.(2010)Zalesky, Fornito, and Bullmore}]{Zalesky2010}
Zalesky, A., Fornito, A., and Bullmore, E.T. (2010).
\newblock Network-based statistic: Identifying differences in brain networks.
\newblock \textit{NeuroImage}, \textbf{53}(4), 1197--1207.

\end{thebibliography}
\bibliographystyle{/home/cgineste/ref/style/oupced3}

% Index -------------------------------------------------------

\addcontentsline{toc}{section}{Index}
%\printindex

\end{document}